\documentclass[preprint,journal]{vgtc}            

\onlineid{0}

\vgtccategory{Research}

\title{A Task-Driven Framework for Multiscale Ocean Flow Dynamics through Integrated Simulation and Visualization}

\author{%
   \authororcid{James Kress}{0000-0002-9706-6182},   
   \authororcid{Jithendra Nadimpalli}{0000-0002-2913-8789},    
   \authororcid{Shehzad Afzal}{0000-0002-9712-3618}, 
   \authororcid{Sohaib Ghani}{0000-0002-9100-7691}, and    
   \authororcid{Ibrahim Hoteit}{0000-0002-3751-4393}
}

\authorfooter{  
  \item  
  	The authors are with King Abdullah University of Science and Technology (KAUST), Thuwal 23955-6900, Saudi Arabia. 
  	E-mail: \{james.kress, jithendra.nadimpalli, shehzad.afzal,  sohaib.ghani, ibrahim.hoteit\}@kaust.edu.sa.  
}

\abstract{%
  Internal waves are large-amplitude gravity waves that occur below the ocean surface and propagate along interfaces separating water layers of different densities. Understanding their generation, propagation, and evolution is essential, as these waves play a vital role in the ocean system by contributing to nutrient transport, biological productivity, and the transfer of energy across the ocean and continental shelf. Domain scientists use high-resolution numerical ocean models, to study internal-wave dynamics and associated coastal and nearshore processes on hybrid computational grids. These models generate large-scale, three-dimensional spatiotemporal datasets that capture internal wave flow behavior and interactions with multiple ocean variables. These datasets are generally analyzed using command-line tools with limited interactivity.
   To address these challenges, we in collaboration with domain scientists designed a task-driven visualization methodology for analyzing multiscale, multivariate flow data on hybrid grids. The framework incorporates a hybrid-grid volumetric reconstruction method, enabling continuous 3D analysis and a coordinated multi-view design that supports interactive exploration of complex flow structures. An insight-based evaluation with domain experts demonstrates that the system enables the identification of previously difficult-to-observe phenomena, including transverse wave propagation, energy transport pathways, and shoaling-driven mixing. Beyond the application domain, our contributions provide generalizable techniques and design principles for visual analysis of multiscale, multivariate flow data on irregular grids.
}


\teaser{
  \centering
  \includegraphics[width=\linewidth, alt={A software dashboard featuring a large central 3D visualization of the ocean floor and volumetric vertical water velocities (colored red and blue), surrounded by several smaller 2D plots and cross-sections showing flow vectors, temperature gradients, and coastal data..}]{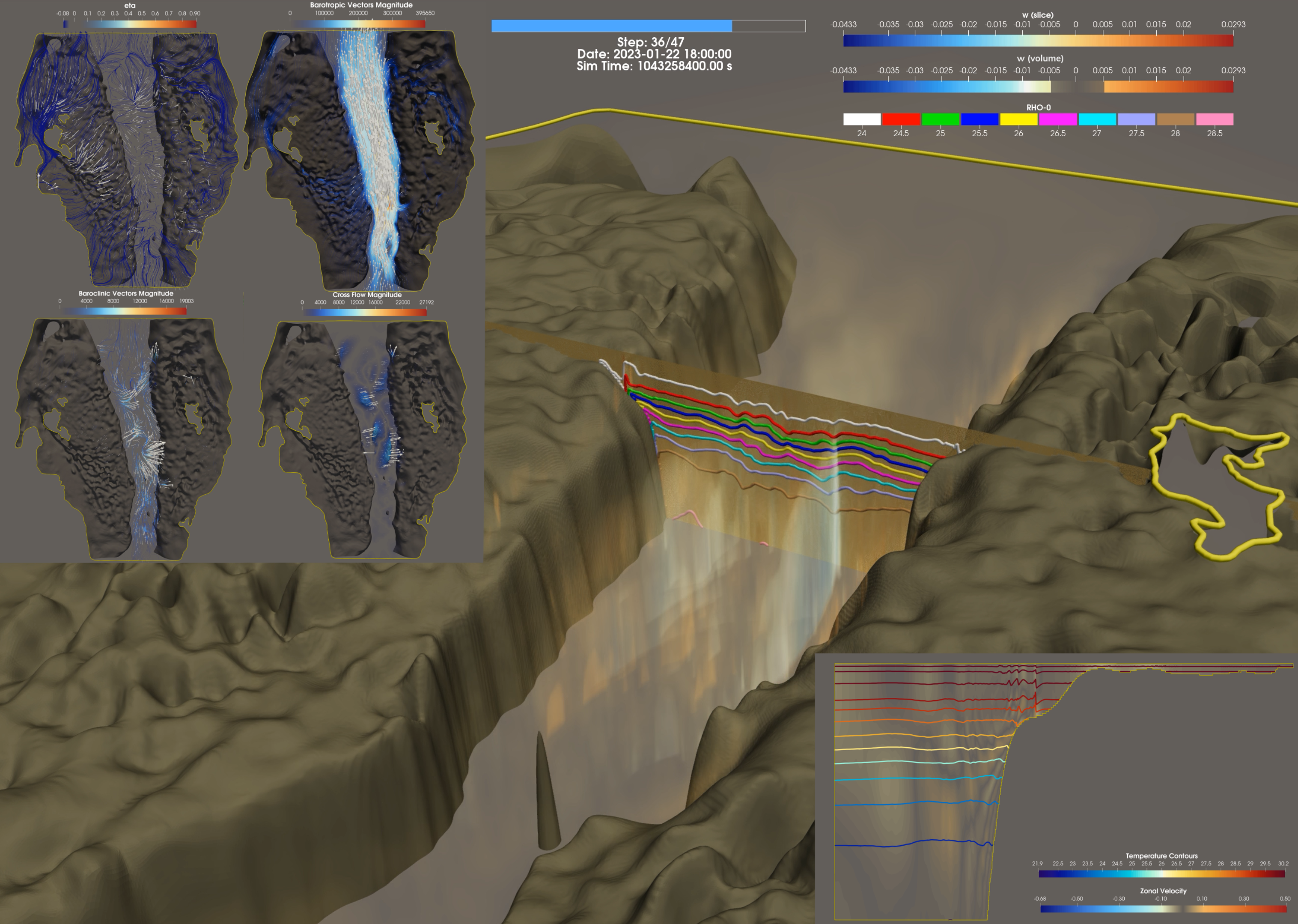}
  \caption{Task-driven multi-view visualization for multiscale flow analysis. The central volumetric rendering highlights 3D flow structures using magnitude-based transfer functions, while surrounding linked views provide complementary context, including surface dynamics, flow decomposition, cross-flow analysis, and multivariate inspection. This coordinated design enables simultaneous exploration of spatial, temporal, and multivariate relationships that are difficult to interpret with traditional slice-based workflows.}
 
  \label{fig:teaser}
}

\graphicspath{{figs/}{figures/}{pictures/}{images/}{./}} 

\usepackage{amsmath}
\usepackage{tabu}                      
\usepackage{booktabs}                  
\usepackage{lipsum}                    
\usepackage{mwe}                       
\usepackage{ccicons}                   
\usepackage{tabularx}
\usepackage{mathptmx}                  
\usepackage{enumitem}
\setlist[itemize]{noitemsep, topsep=2pt, leftmargin=*}

\begin{document}

\firstsection{Introduction}

\maketitle

Nonlinear internal waves (NLIWs)---key components of oceanic flows---are gravity waves that propagate below the ocean surface along strong density gradients. Understanding their generation, propagation, and dissipation remains an active area of research. Their dynamics are governed by multiple factors, including stratification, tidal forcing, complex underwater topography, and wave--wave interactions~\cite{alford2015formation}. NLIWs play a central role in ocean processes such as vertical mixing, nutrient transport, sediment transport, energy redistribution across ocean basins, biological productivity, and ecological variability. As these waves approach shallow regions, they shoal and dissipate, delivering cooling and nutrient-rich waters onto the shelf, which can benefit coral ecosystems and influence biological activity~\cite{muller1986nonlinear}. Domain scientists use numerical simulation models such as SUNTANS (Stanford Unstructured Nonhydrostatic Terrain-following Adaptive Navier--Stokes Simulator)~\cite{fringer2006unstructured} to capture the complex flows governing the evolution of NLIWs in the presence of intricate coastlines, variable bathymetry, and shallow regions. Analyzing these large-scale simulations requires methods for exploring multiscale, multivariate, and time-varying data defined on hybrid computational grids. Such datasets arise broadly in oceanographic simulations, where unstructured grids are essential for accurately representing coastal and topographically complex regions. Although advances in numerical modeling have enabled high-fidelity simulation of these systems, extracting meaningful insight from the resulting data remains challenging. 

A primary challenge arises from how the data is structured. Many ocean simulations, particularly those focused on phenomena such as NLIWs, are defined on hybrid computational grids that combine horizontally unstructured meshes with vertically varying terrain-following layers. While this representation is essential for capturing complex ocean and coastal geometries, it lacks the explicit volumetric connectivity required for continuous spatial analysis. Consequently, domain scientists often rely on slice-based workflows involving manual inspection of multiple 2D transects across space and time. This approach fragments spatial context, limits multivariate reasoning, and makes it difficult to track the evolution of 3D structures. 

Beyond representational issues, analysis is further complicated by interactions among multiscale flow components. In-depth study of such flow requires separating dominant large-scale transport from smaller-scale dynamics that may also be scientifically important. However, standard visualization approaches provide limited support for isolating these interacting components, forcing domain scientists to rely on indirect and iterative methods. In addition, understanding complex oceanic processes requires examining multivariate correlations across space and time, which conventional workflows support inadequately.

The challenges posed by oceanic flow datasets such as internal wave simulations~\cite{xie2019survey, Lv2022_multidim_visualization_ocean_data, Zang2023_internal_wave_sediment_transport} reflect broader issues in analyzing multiscale flow data on irregular grids. These include the need for continuous volumetric representations, context-aware visualization, feature identification and tracking in stratified oceans, separation of cross-scale flow interactions, characterization of energy transport pathways, and coordinated multivariate exploration. 
In this work, we present a task-driven visualization methodology for analyzing multiscale, multivariate flow data defined on hybrid computational grids. 
 
{\color{black}A key feature of our approach is a cross-flow decomposition capability that uses the locally dominant flow direction as a reference to isolate transverse wave-related motion, revealing multiscale flow dynamics more clearly.}
Combined with a volumetric reconstruction pipeline and coordinated multiview design, the framework enables efficient exploration of complex flow phenomena. 

We demonstrate its effectiveness through insight-driven evaluation with domain experts, showing that it reveals structures difficult to observe with traditional slice-based workflows while reducing analysis effort.
Although motivated by oceanographic applications, our methods operate on scalar and vector fields defined on irregular grids and are applicable to a broad class of flow analysis problems. The main contributions of our work are as follows: 

\begin{enumerate}
    \item \textbf{Hybrid-Grid Volumetric Reconstruction}: A data transformation pipeline that converts hybrid simulation outputs into a cohesive 3D unstructured volume with explicit vertical connectivity and cell-to-point interpolation, enabling continuous volumetric analysis beyond conventional slice-based workflows.
            
    \item \textbf{Cross-Flow Decomposition}: 
    {\color{black}A visualization-oriented decomposition that uses the locally dominant flow direction as a reference to separate a flow-aligned component from a transverse residual component. The resulting cross-flow field enhances the visual detection of transverse internal-wave propagation and lateral energy pathways that are otherwise obscured by strong background flow.}

    \item \textbf{Multi-View Design Framework}: A coordinated visualization design integrating volumetric rendering, flow decomposition, and interactive slicing, feature highlighting, and multivariate analysis applicable to large-scale 3D flow data.
    
    \item \textbf{Insight-Driven Evaluation and Case Studies}: A set of case studies and feedback from domain experts demonstrating in-depth interactive analysis capabilities, including the identification of difficult-to-observe structures, energy pathways, and multiscale flow interactions, compared to traditional workflows.  
    
\end{enumerate}

\section{Background}

Internal waves are gravity-driven oscillations that propagate along density interfaces within stratified layers. Generated by variations in temperature and salinity, they play a critical role in ocean dynamics, including vertical mixing, nutrient transport, and energy redistribution. As they interact with complex bathymetry and coastal regions, internal waves can undergo nonlinear steepening, propagation, and dissipation, influencing both ecological and physical marine processes~\cite{grimshaw2010internal}.

Several mechanisms have been identified globally for exciting these disturbances~\cite{maxworthy1979note, da2015internal, bourgault2016generation, raju2021numerical}, among which tide--topography interaction is the most dominant. These nonlinear waves can travel long distances before dissipating, thereby contributing to vertical mixing and transporting energy across the ocean~\cite{moum2007energy, mathur2025internal}. They also induce strong velocities that vertically displace isopycnals {\color{black}(surfaces of constant density)}~\cite{nadimpalli2025interaction}, influencing nutrient transport, biological productivity~\cite{pan2012enhancement, muacho2013effect}, offshore engineering structures~\cite{song2011comparisons, wang2018numerical}, underwater acoustics~\cite{apel2007internal}, and submarine operations~\cite{he2022numerical}. Studying the generation, propagation, and dissipation of large-amplitude NLIWs requires very high-resolution spatiotemporal ocean models such as SUNTANS~\cite{fringer2006unstructured}. High resolution is necessary to accurately represent tide--bathymetry interactions, resolve the balance between nonlinear steepening and nonhydrostatic dispersion, and capture wave--wave interactions and dissipation processes across scales.

These ocean models simulate fluid flow on hybrid computational grids that combine horizontally unstructured polygonal meshes with vertically varying terrain-following layers to accurately represent complex bathymetry and coastal geometries. While this hybrid structure enables high-fidelity simulation, it introduces substantial visualization challenges. The data is inherently multivariate, time-varying, and defined on irregular geometries, with only implicit vertical connectivity between layers. As a result, standard visualization pipelines cannot directly represent these datasets as continuous volumes, limiting the applicability of interpolation-based techniques such as volume rendering and streamline computation. The scale of these simulations further compounds the problem, as datasets often contain millions of grid elements and multiple physical variables (e.g., velocity components, temperature, salinity, and density) evolving over time. This creates a need for an efficient visualization framework capable of handling large data volumes while supporting analysis of multiple variables across space and time.

To analyze these simulation outputs, the oceanography community commonly relies on generic tools such as MATLAB, Matplotlib, ArcGIS, Tecplot, etc.
While these tools provide flexible data access and basic visualization capabilities, they exhibit several limitations for analyzing hybrid-grid, multivariate flow data:
\begin{itemize}
    \item \textbf{Fragmented spatial representation:} Analysis is typically performed using disconnected 2D slices or transects, limiting the ability to reason about continuous 3D structures.
    
    \item \textbf{Limited support for hybrid grids:} Standard tools assume structured or uniformly unstructured grids, making it difficult to represent terrain-following vertical layers as cohesive volumes.
    
    \item \textbf{Inadequate flow separation:} Dominant large-scale flows often obscure smaller-scale dynamics, and existing methods provide limited support for isolating interacting flow components.
    
    \item \textbf{Lack of coordinated multivariate exploration:} Variables are often analyzed independently, requiring manual alignment across plots, time steps, and spatial regions.
\end{itemize}
These limitations result in iterative, time-consuming workflows that place a high cognitive burden on analysts and hinder effective exploration of complex spatiotemporal phenomena.

In the atmospheric and earth sciences domain, interactive analysis of multidimensional, multivariate, and complex observational and modeled data is crucial for advanced analysis and decision-making. A variety of tools have been developed to support such exploration~\cite{rautenhaus2017visualization, afzal2019state, xie2019survey}. However, for analysis of internal wave flow, we could not identify any custom interactive tool that fully meets domain scientists' needs. Most prior work relies on standard tools either to detect internal waves or to analyze the associated complex datasets~\cite{internalwaves, dematteis2024interacting, sutherland1999visualization, rodenas1998new, VASAVI2021145, chashechkin2011numerical, WHITFORD2002537, Jain2023_pyParaOcean, afzal2019redseaatlas}. There is also work on animating internal wave flow using generic tools such as MATLAB~\cite{gould2000computer,hinsinger2002interactive}. Although general-purpose visualization tools such as VAPOR~\cite{clyne2007interactive}, pyFerret~\cite{hankin1992ferret}, and ParaView~\cite{ahrens2005paraview} are often used to analyze large oceanographic datasets, they typically assume standard structured or unstructured grids. These tools lack native support for the specialized hybrid topology used in ocean models, which combines horizontally unstructured polygons with vertically structured or unstructured layers to capture the undulations of the ocean floor. Because standard tools cannot natively resolve these varying vertical geometries, domain scientists are often forced to rely on disconnected 2D slices or basic scripting tools. This structural incompatibility highlights a critical research gap and motivates the custom data processing workflow developed in this study. 

\begin{figure}[htb]
    \centering
    \includegraphics[width=1\linewidth]{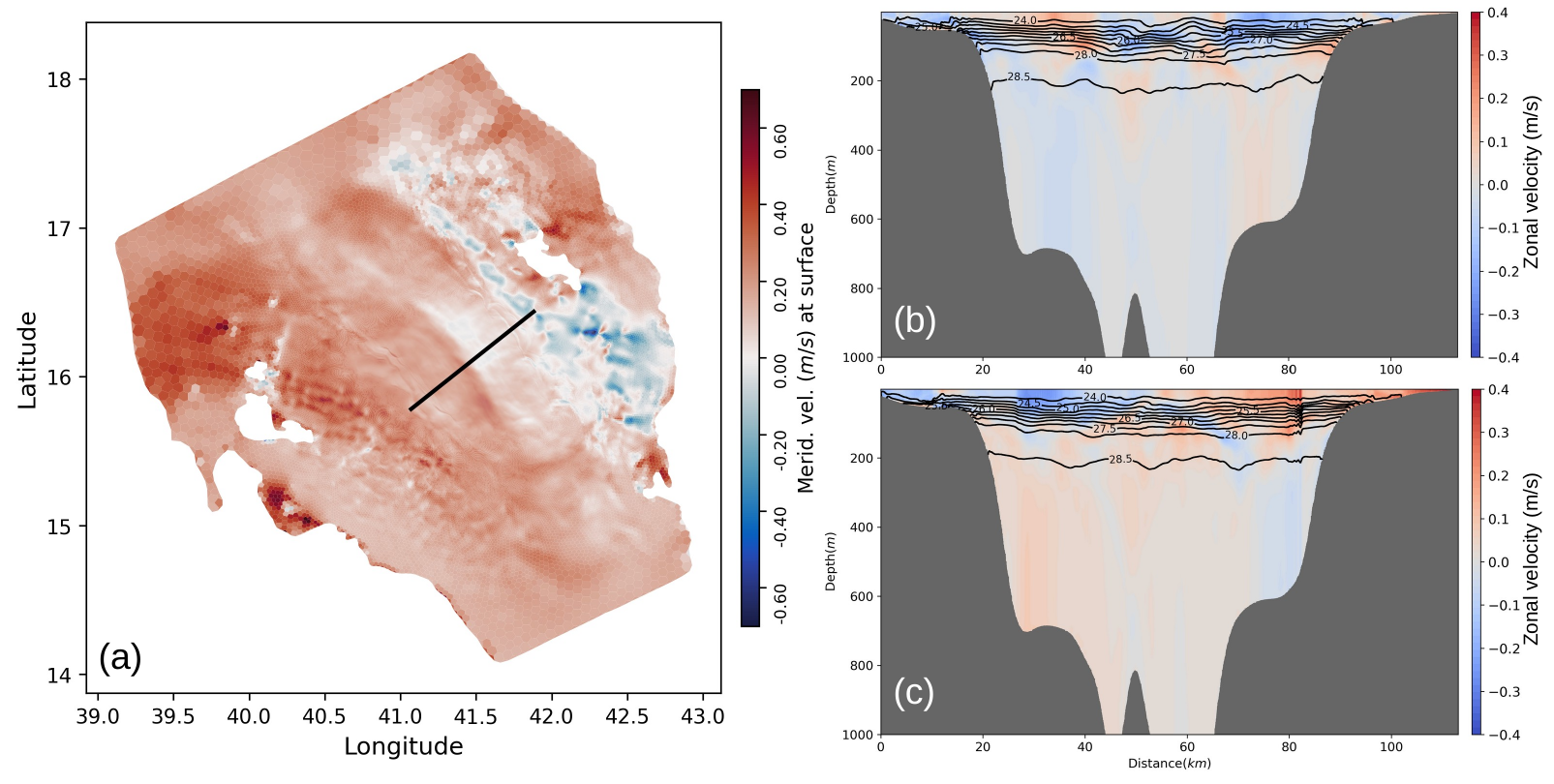}
    \caption{Conventional slice-based analysis limitations.(a) Spatial distribution of surface velocity on an unstructured grid. (b–c) Repeated 2D transects (black line(a)) required to analyze spatiotemporal evolution.}     
    \label{fig:2Dslice}
\end{figure}

To further illustrate a typical domain-scientist workflow, consider the example of visualizing output from an unstructured-grid ocean model. Fig.~\ref{fig:2Dslice}(a) shows a spatial plot of surface velocity for a single time step, generated using the Matplotlib package in Python. Individual grid cells are combined to form polygons, which are then colored according to normalized velocity values. Rendering such polygon-based visualizations is computationally expensive even on moderately capable systems. These plots must be repeatedly examined across time steps, depths, and variables to understand the spatial dynamics of internal waves. In addition, 2D vertical slices along the black line in Fig.~\ref{fig:2Dslice}(a) are generated for density and zonal velocity in the x--z plane at different times to track the propagation of NLIWs, which appear as sharp density interfaces in Figs.~\ref{fig:2Dslice}(b--c). This repetitive workflow is tedious and time-consuming. A visualization tool that enables interactive extraction of spatial slices, variables, and time steps would therefore improve the efficiency and effectiveness of oceanographic data analysis.

\vspace{-8pt}

\section{Motivation and Analytical Tasks}
\label{sec:designreq}

This work is a collaboration between ocean and atmospheric scientists and visualization researchers. Two coauthors bring extensive experience in physical oceanography, numerical modeling, and data assimilation. Through a series of exploratory sessions, we identified key challenges in analyzing simulation data and discussed interactive solutions. 
In the initial phase, domain experts focused on analyzing the evolution, propagation, and dynamics of oceanic flows. Their goals included examining wave--coastline interactions and impacts, assessing the vertical displacement of {\color{black}isopycnals} within user-defined cross-sections, evaluating energy transport from deep to shallow regions, analyzing temporal variations in temperature, salinity, and related variables, understanding the influence of bathymetry on flow dynamics, and studying nutrient and sediment transport toward coastal areas. 

These analysis capabilities can support visual hypothesis validation and decision-making in applications such as offshore infrastructure, coastal ecosystem health, ocean circulation and mixing, submarine operations, acoustic communications, and model validation.
Experts reported that their current workflows rely mainly on 2D slices for selected parameter combinations, making 3D variable interactions difficult to interpret.
They wanted an interactive 3D environment with multiple views that could present information aligned with their analysis requirements in a cohesive manner. 
They also needed data-processing workflows capable of loading newly generated simulation runs in a high-performance computing environment. 
After multiple iterations, we distilled these domain challenges into the following key analytical tasks to guide the design of our 3D visualization workflow. These abstractions generalize to broader classes of multiscale, multivariate flow analysis problems.

\begin{itemize}

\item \textbf{T1: Data Transformation and Representation.} 
Transform hybrid, multiresolution simulation outputs into a continuous volumetric representation that supports coherent spatial analysis across irregular geometries and varying vertical structures.

\item \textbf{T2: Spatiotemporal Contextualization.} 
Provide persistent geospatial context, including bathymetry and coastline features, to support the interpretation of flow--topography interactions and spatially localized phenomena.

\item \textbf{T3: Feature Identification and Spatiotemporal Tracking.} 
Identify and track salient structures in temporal 3D scalar and vector fields, enabling analysis of their evolution, propagation, and deformation.

\item \textbf{T4: Vector Field Decomposition and Analysis.} 
Separate and analyze interacting components within multivariate vector fields to reveal mechanisms that are not directly observable in raw data.

\item \textbf{T5: Multivariate Exploration and Quantitative Inspection.} 
Support interactive, user-driven exploration through synchronized, multi-view probing of thermodynamic properties (temperature, salinity, density) across arbitrary cross-sections of the 3D domain.

\end{itemize}


\section{{\color{black}Visualization Framework for Ocean Flow Dynamics}}

\begin{figure}[htbp]
    \centering
    \includegraphics[width=\linewidth]{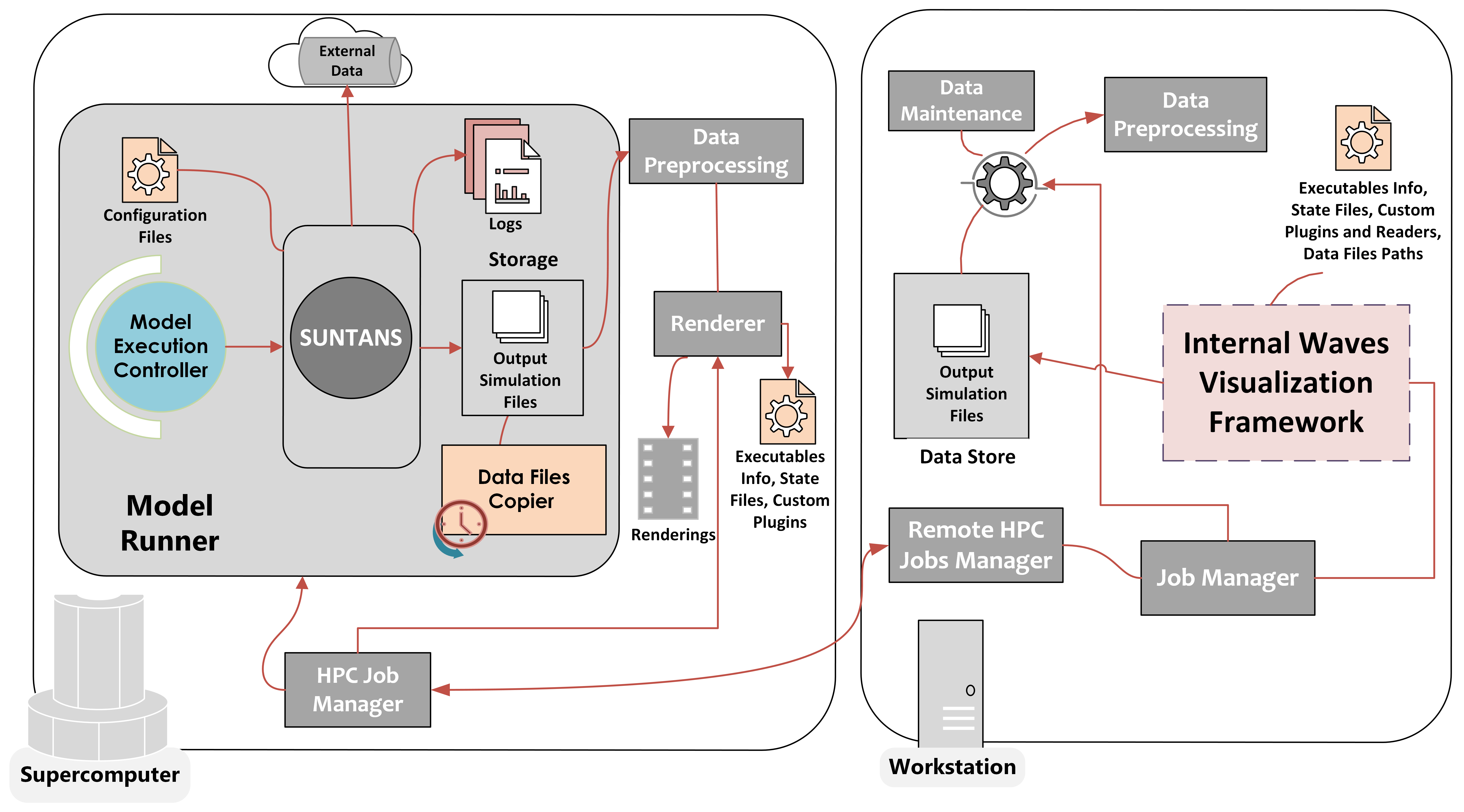}
    \caption{System architecture for hybrid-grid flow analysis. Simulation data from HPC is processed via a pipeline for interactive, scalable visualization across local and distributed environments.    
    }
    \label{fig:system_arch}
\end{figure}

Fig.~\ref{fig:system_arch} shows the architecture of our ocean flow visualization framework. While demonstrated with SUNTANS~\cite{fringer2006unstructured}, the design generalizes to other models. The HPC-centric workflow supports large-scale simulations, though the framework is hardware-agnostic for 
local workflows. 
The model runner manages supercomputer execution and in situ processing, while the Job Manager coordinates scheduling, data synchronization, and communication between HPC and local environments.

\subsection{Data and Ocean Model Simulation Context} 

Our analysis uses high-resolution SUNTANS simulations~\cite{fringer2006unstructured} of multiscale flow dynamics in stratified environments. The simulations generate time-varying multivariate fields on hybrid grids, including velocity, temperature, salinity, density, and sea surface height. We focus on NLIWs in the southern Red Sea, where tide–topography interactions generate large-amplitude waves that propagate across the basin and dissipate near coastal regions. Prior studies identify key baroclinic tidal energy generation regions, including the Bab al-Mandeb Strait, southern Red Sea shelf breaks, Gulf of Suez, and Strait of Tiran~\cite{guo2018baroclinic}, with observations~\cite{da2012internal} and simulations~\cite{guo2016generation} showing the evolution of these waves into internal solitary-like waves.

To simulate NLIWs, the nonlinear, non-hydrostatic SUNTANS configuration~\cite{fringer2006unstructured} is applied over $13^\circ$–$19^\circ$N, solving Reynolds-averaged Navier–Stokes equations under the Boussinesq approximation on horizontally unstructured $n$-sided polygons with a vertically structured grid. Hexagons dominate ($\approx$96.8\%), reducing divergence errors~\cite{holleman2013numerical}. Horizontal resolution is $\approx$100 m in wave-generation regions, coarsening to 10 km near open boundaries. Bathymetry is from GEBCO~\cite{gebco2022}, initial temperature and salinity are from the World Ocean Atlas 2023~\cite{woa2023}, and open boundaries forced with OTIS TPXO9 barotropic tides~\cite{egbert_erofeeva_2002,tpxo9}. Simulations span 15 days, covering spring and neap tides, executed on 240 Shaheen~III processors~\cite{shaheen3} with ParMETIS~\cite{karypis2011metis}. Outputs include 3D scalar and vector fields and sea surface height in NetCDF4.

The dataset’s high resolution, irregular geometry, hybrid vertical structure, and temporal variability motivate our task-driven design: hybrid-grid reconstruction for volumetric representation (T1), context-aware geospatial visualization (T2), and coordinated multivariate analysis for feature identification and decomposition (T3–T5). Conventional slice-based workflows cannot capture coherent 3D structures, cross-variable relationships, or temporal evolution, highlighting the broader relevance of this data class to multiscale simulations on irregular grids.


\subsection{Hybrid-Grid Volumetric Reconstruction}
\label{sec:suntans-reader}
\begin{figure}[t!]
    \centering 
    \begin{subfigure}{1\columnwidth}
        \centering
        \includegraphics[width=\linewidth]{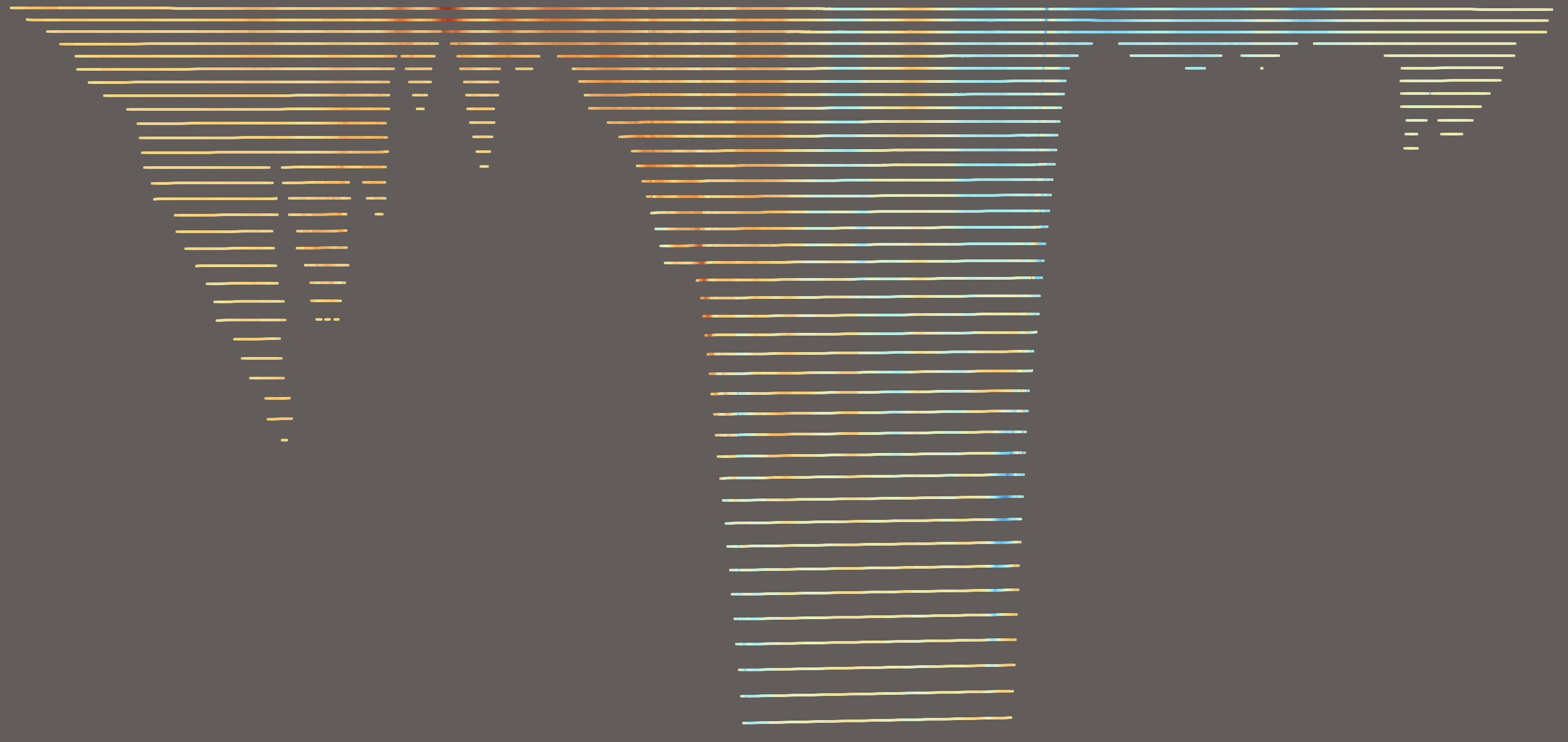}        
        \caption{Initial data transformation state: Model data rendered as disconnected 2D slices with uniform vertical offsets, preventing volumetric interpolation.}
        \label{fig:unconnected_slices}
    \end{subfigure}
    \hfill 
    \begin{subfigure}{1\columnwidth}
        \centering        
        \includegraphics[width=\linewidth]{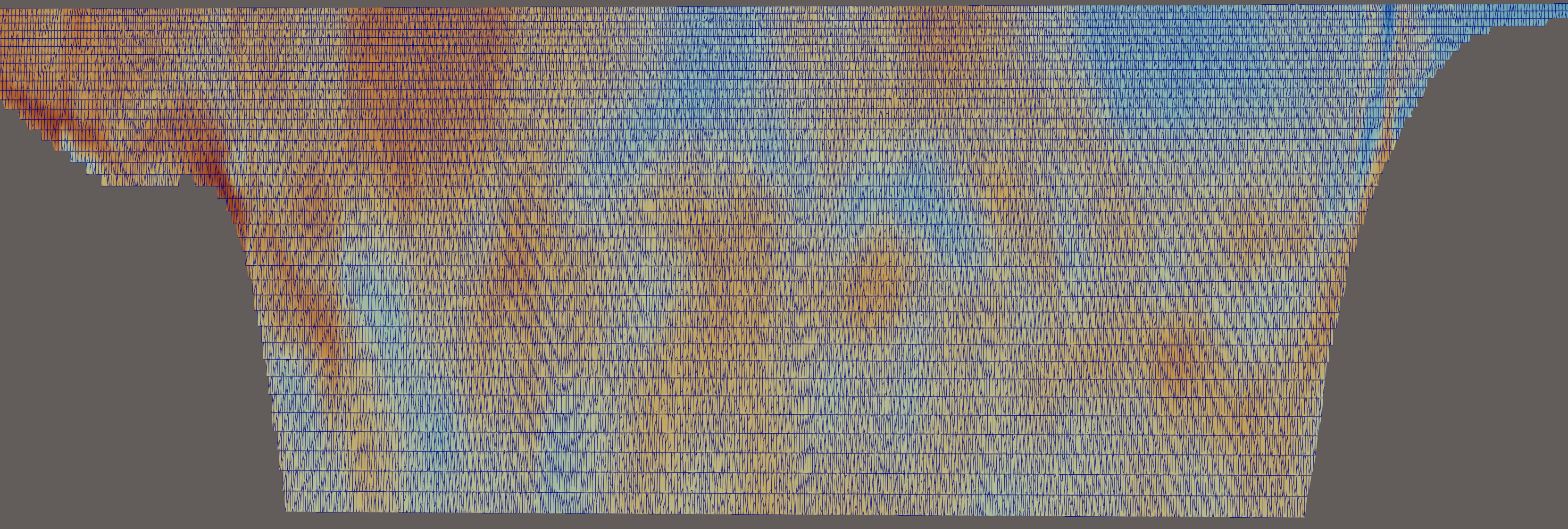}
        \caption{Final data transformation state: Reconstructed cohesive 3D volume with explicit vertical connectivity and correct depth coordinates.}
        \label{fig:vertically_connected_slices}
    \end{subfigure}
    \hfill 
    \begin{subfigure}{.48\columnwidth}
        \centering
        \includegraphics[width=\linewidth]{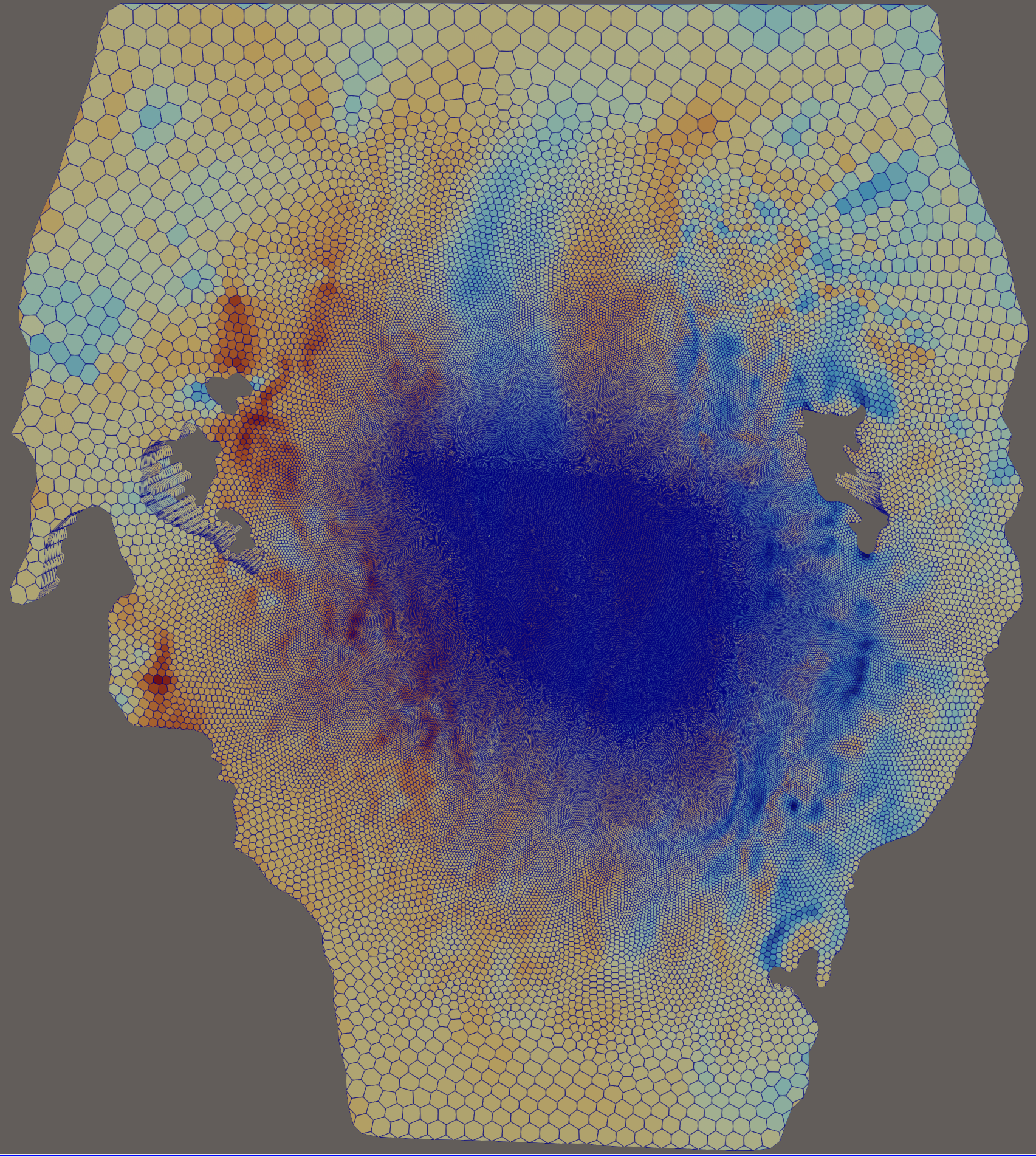}        
        \caption{{\color{black}Initial state: Traditional cell-centered data distribution; note the lack of continuity at slice boundaries.}}
       \label{fig:cell_centered}
    \end{subfigure}
    \hfill
    \begin{subfigure}{.48\columnwidth}
        \centering
        \includegraphics[width=\linewidth]{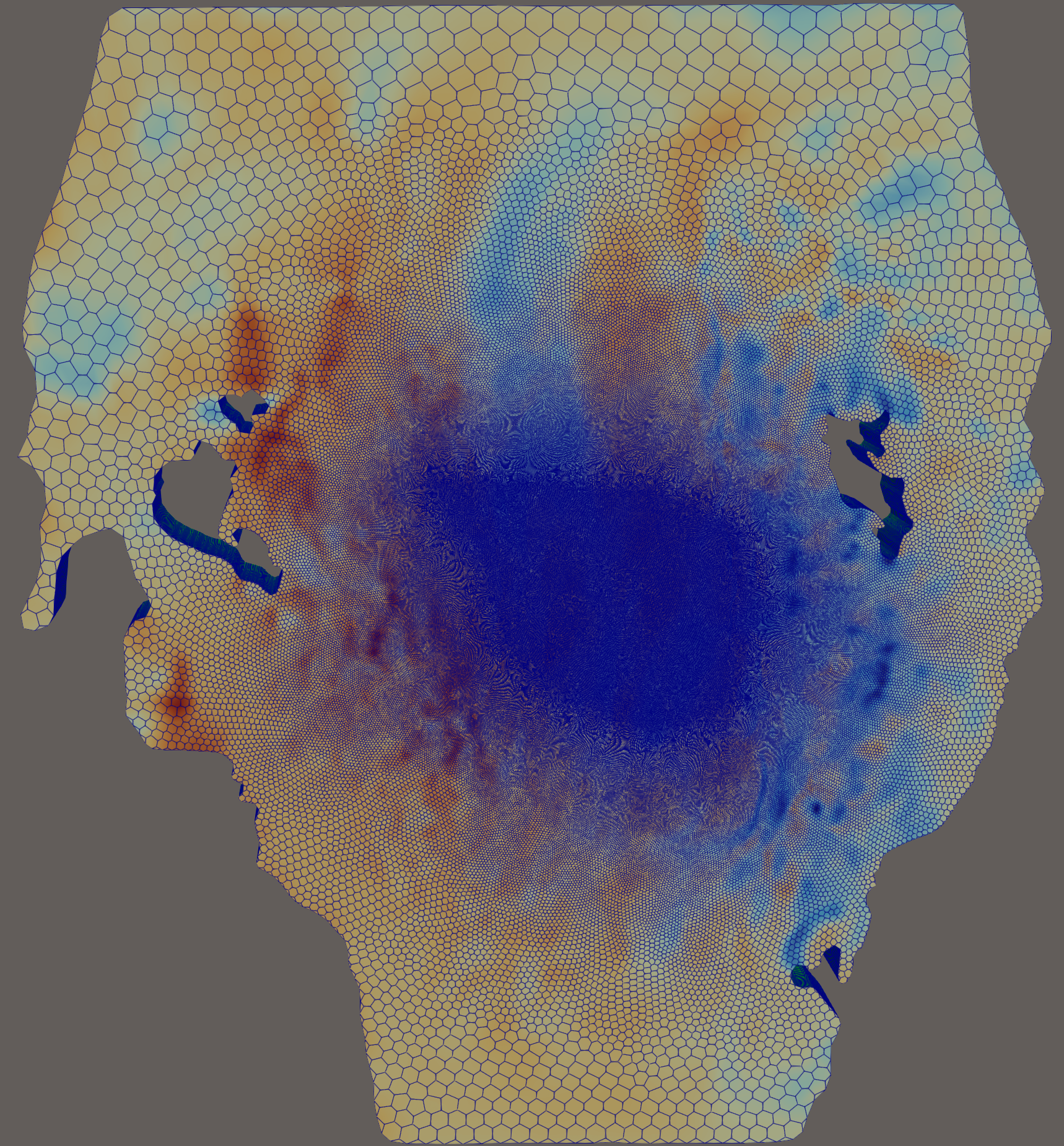}      
        \caption{Final state: Point-centered data on a unified grid, enabling high-fidelity 3D rendering and gradient analysis.}             
        \label{fig:point_centered}
    \end{subfigure}
    \hfill 
    \begin{subfigure}{1\columnwidth}
        \centering
        \includegraphics[width=\linewidth]{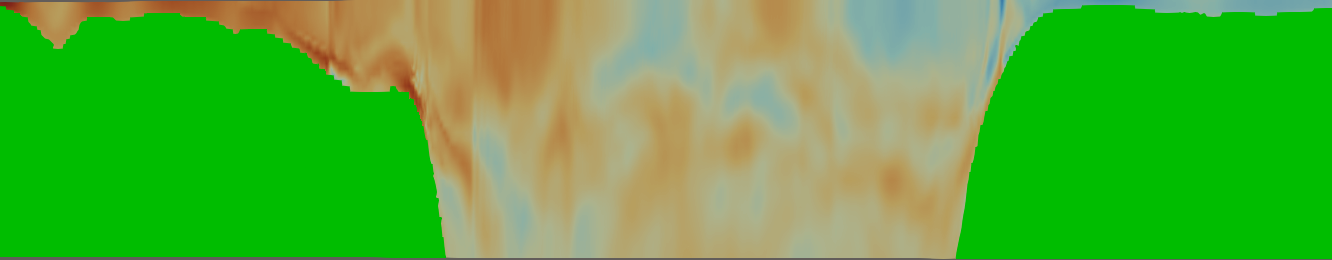}
         \caption{Optimized handling of inactive grid cells; mapping simulated regions (colored) and excluding NaN values (green).}
        \label{fig:nan_values}
    \end{subfigure}
    \caption{Evolution of the Hybrid-grid volumetric reconstruction. The framework moved from discrete 2D slices (Figs.~\ref{fig:unconnected_slices}, \ref{fig:cell_centered}) to a unified, point-centered 3D unstructured grid (Figs.~\ref{fig:vertically_connected_slices}, \ref{fig:point_centered}). This development enables volumetric analysis and 3D visualization of the ocean flows}    
    \label{fig:suntans-paraview-reader}
\end{figure}

The custom NetCDF output format and highly specialized hybrid grid structure of SUNTANS presented a significant challenge for visualization with conventional tools. The raw data, which utilizes hexagon-dominated cells horizontally and a structured, terrain-following arrangement vertically, was unreadable by standard ParaView\cite{ahrens2005paraview} plugins and generic NetCDF readers due to its custom mesh topology. This necessitated the development of a specialized reader. In its initial implementation, the custom reader successfully loaded the mesh coordinates but rendered the domain as distinct, two-dimensional horizontal slices with uniform vertical spacing, failing to reflect the non-uniform depth of the terrain-following layers (See Fig.~\ref{fig:unconnected_slices}). This absence of explicit vertical connectivity and correct spacing was a critical limitation, resulting in a disconnected layer structure that prevented the use of volumetric techniques essential for 3D dynamics analysis, such as streamline generation and isosurfacing.

To address this limitation, we reconstruct the dataset as a single 3D unstructured volume by explicitly establishing vertical connectivity between layers. The reader connects corresponding nodes across adjacent layers and uses the model-provided depth coordinates (e.g., $\text{z\_r}$ and $\text{z\_w}$) to determine their precise vertical positions. This process preserves the terrain-following structure and transforms the layered representation into a geometrically consistent volume by forming 3D wedges and hexahedral prisms from the horizontal mesh elements (Fig.~\ref{fig:vertically_connected_slices}). This reconstruction directly addresses T1 by enabling a continuous volumetric representation of hybrid-grid simulation data.

To ensure numerical consistency and enable continuous analysis, all cell-centered variables (e.g., $\text{uc}$, $\text{vc}$, $\text{salt}$, $\text{temp}$) are interpolated to mesh nodes, producing point-centered data (Figs.~\ref{fig:cell_centered} and~\ref{fig:point_centered}). This transformation aligns variable values with the shared vertices of the reconstructed elements, preventing data loss at layer boundaries and enabling stable gradient and derived quantity computation. Regions outside the active simulation domain are explicitly marked as NaN, to exclude them from rendering and analysis (Fig.~\ref{fig:nan_values}).

Additional handling is incorporated for edge-based variables (e.g., edge-normal velocity $\text{U}$), which are exported as line elements with interpolated point data. The reader also reconciles the dual vertical coordinate systems ($\text{z\_r}$ and $\text{z\_w}$) used by different variables, integrating them into a unified representation that correctly positions the full velocity field ($\text{uc}, \text{vc}, \text{w}$) within the volume. Together, these steps enable faithful reconstruction of the simulation domain as a continuous 3D field suitable for volumetric analysis.

The reader is implemented as a custom ParaView plugin and is inherently platform-independent, supporting deployment across local workstations and HPC environments. Integrated into our distributed workflow, it processes large SUNTANS (or other ocean models) outputs directly on the Shaheen~III supercomputer~\cite{shaheen3}, enabling scalable, in situ--compatible post-processing of high-resolution ocean simulations.

\subsection{Visualization Design}
\label{sec:vis_design}

\textbf{Design Rationale}: 

Our design leverages task abstractions (T1–T5) to translate domain requirements (Section~\ref{sec:designreq}) into a coordinated multiview framework for multiscale, multivariate ocean flow analysis on unstructured hybrid grids. For T1 (Data Transformation and Representation), we build a cohesive 3D volumetric representation with explicit vertical connectivity, point-centered interpolation, and unstructured horizontal grids, enabling high-fidelity rendering of fine-scale ocean structures. 
For T2 (Spatial Contextualization), bathymetry and coastline geometry provide geospatial context for interpreting flow–topography interactions and dominant/transverse dynamics.
For T3 (Feature Identification and Spatiotemporal Tracking), magnitude-based volumetric rendering suppresses low-signal regions while highlighting salient structures; synchronized views and temporal navigation support tracking of evolution, propagation, and deformation. For T4 (Vector Field Decomposition and Analysis), interacting flow components are isolated by scale and visualized with streamlines and glyphs using magnitude-driven opacity, reducing clutter and revealing dominant and transverse patterns with consistent encodings for comparative analysis. For T5 (Multivariate Exploration and Quantitative Inspection), linked views and interactive slicing enable correlated analysis across variables and arbitrary cross-sections. Across all tasks, the design emphasizes (i) magnitude-based filtering, (ii) feature-focused encoding, and (iii) tightly coupled multi-view coordination for efficient multivariate reasoning.

\begin{figure*}[t]
\centering

\begin{subfigure}[t]{0.240\textwidth}
    \centering
    \includegraphics[width=\linewidth]{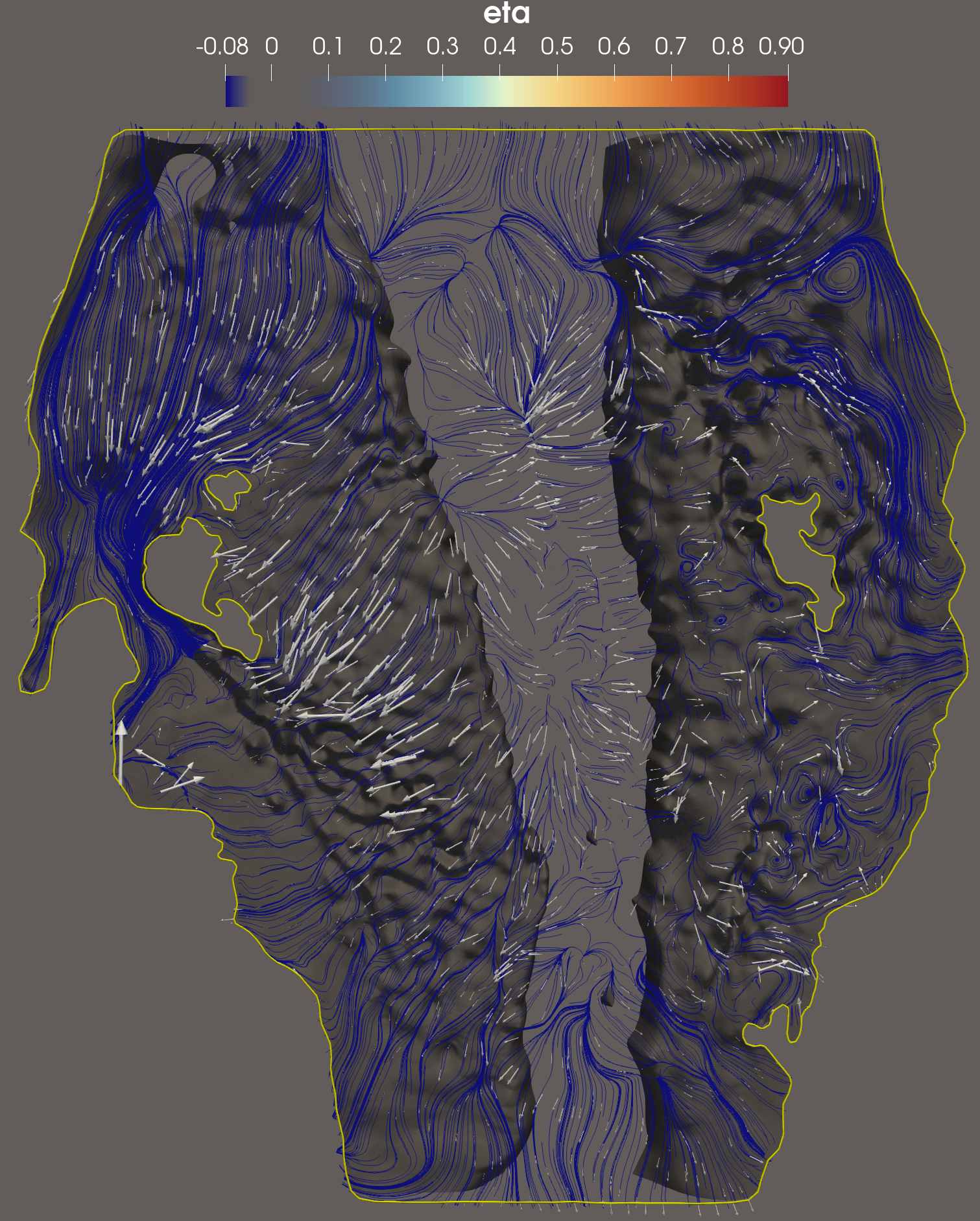}
    \caption{Sea surface elevation (\texttt{eta}) visualized with transparent streamlines and vector glyphs to highlight surface tidal advection.}
    \label{fig:eta}
\end{subfigure}
\hfill
\begin{subfigure}[t]{0.242\textwidth}
    \centering
    \includegraphics[width=\linewidth]{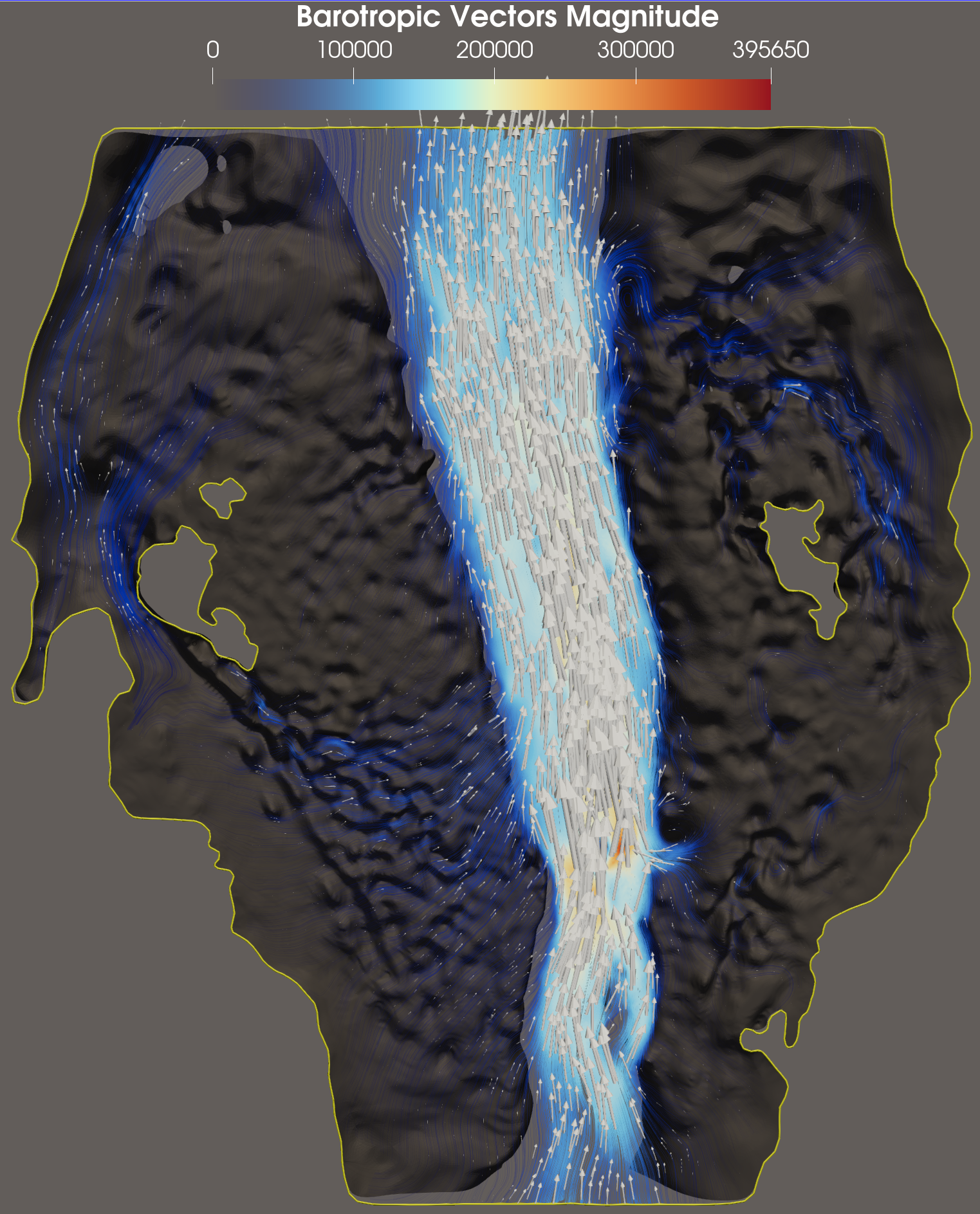}
    \caption{Depth-averaged barotropic flow vectors, illustrating the bulk motion of the water column driven primarily by tidal forcing.}
    \label{fig:barot}
\end{subfigure}
\hfill
\begin{subfigure}[t]{0.245\textwidth}
    \centering
    \includegraphics[width=\linewidth]{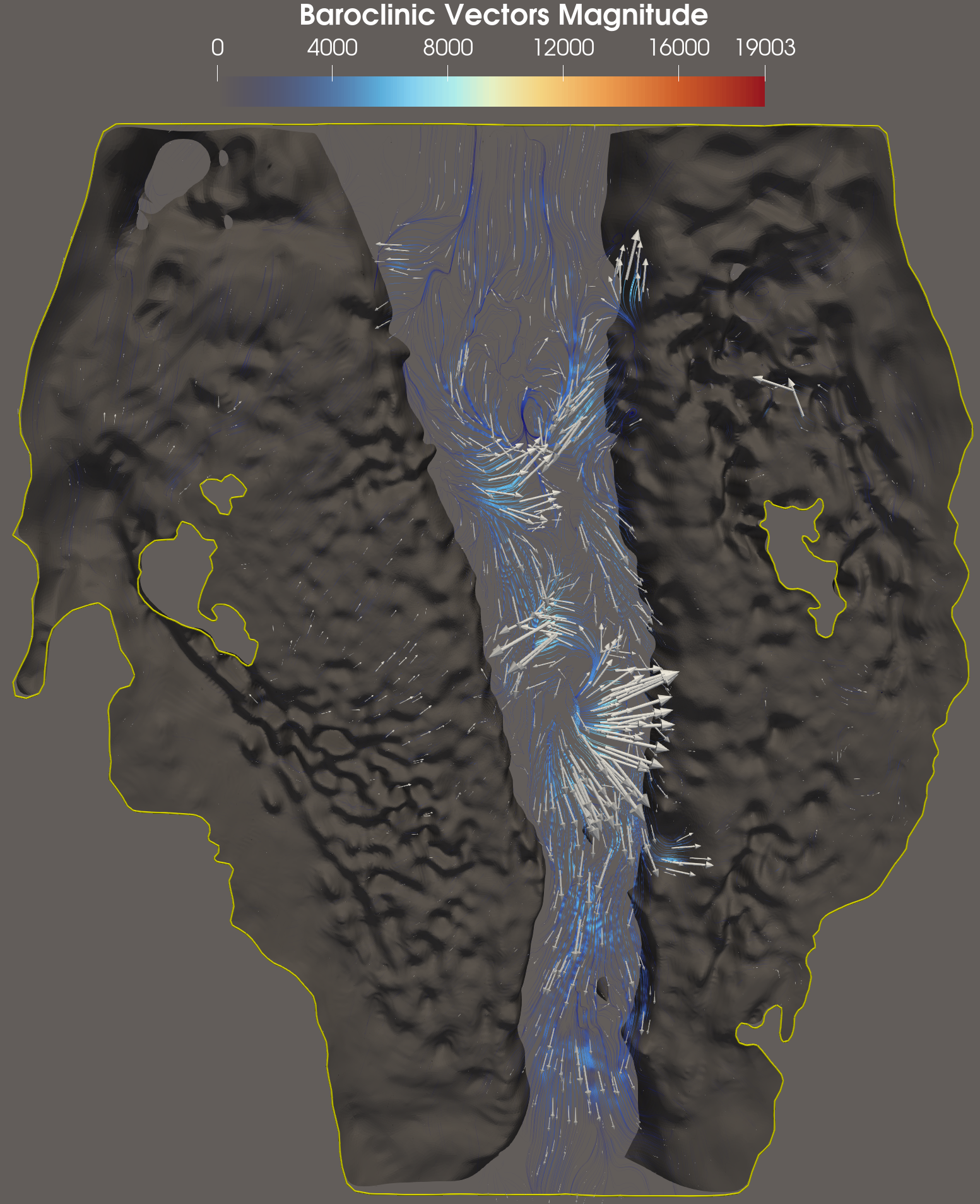}
    \caption{Density-driven baroclinic flux vectors, revealing vertical shear, internal wave beams, and localized energy transport.}
    \label{fig:baroc}
\end{subfigure}
\hfill
\begin{subfigure}[t]{0.24\textwidth}
    \centering
    \includegraphics[width=\linewidth]{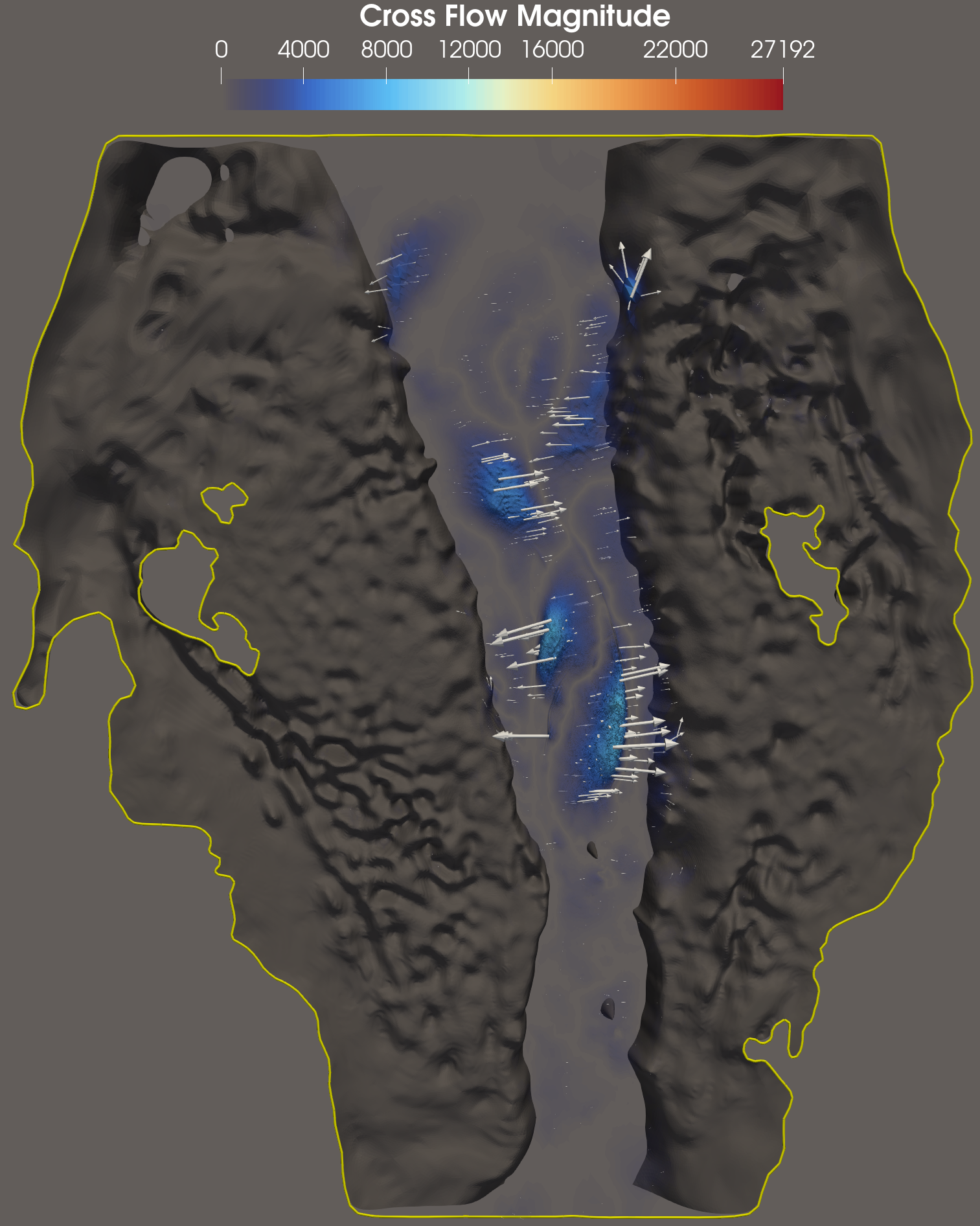}
    \caption{Derived \texttt{cross-flow} vector field, mathematically isolating transverse waves by filtering out dominant barotropic advection.}
    \label{fig:cross_flow}
\end{subfigure}
\vspace{0.5em}
\begin{subfigure}[t]{0.495\textwidth}
    \centering
    \includegraphics[width=\linewidth]{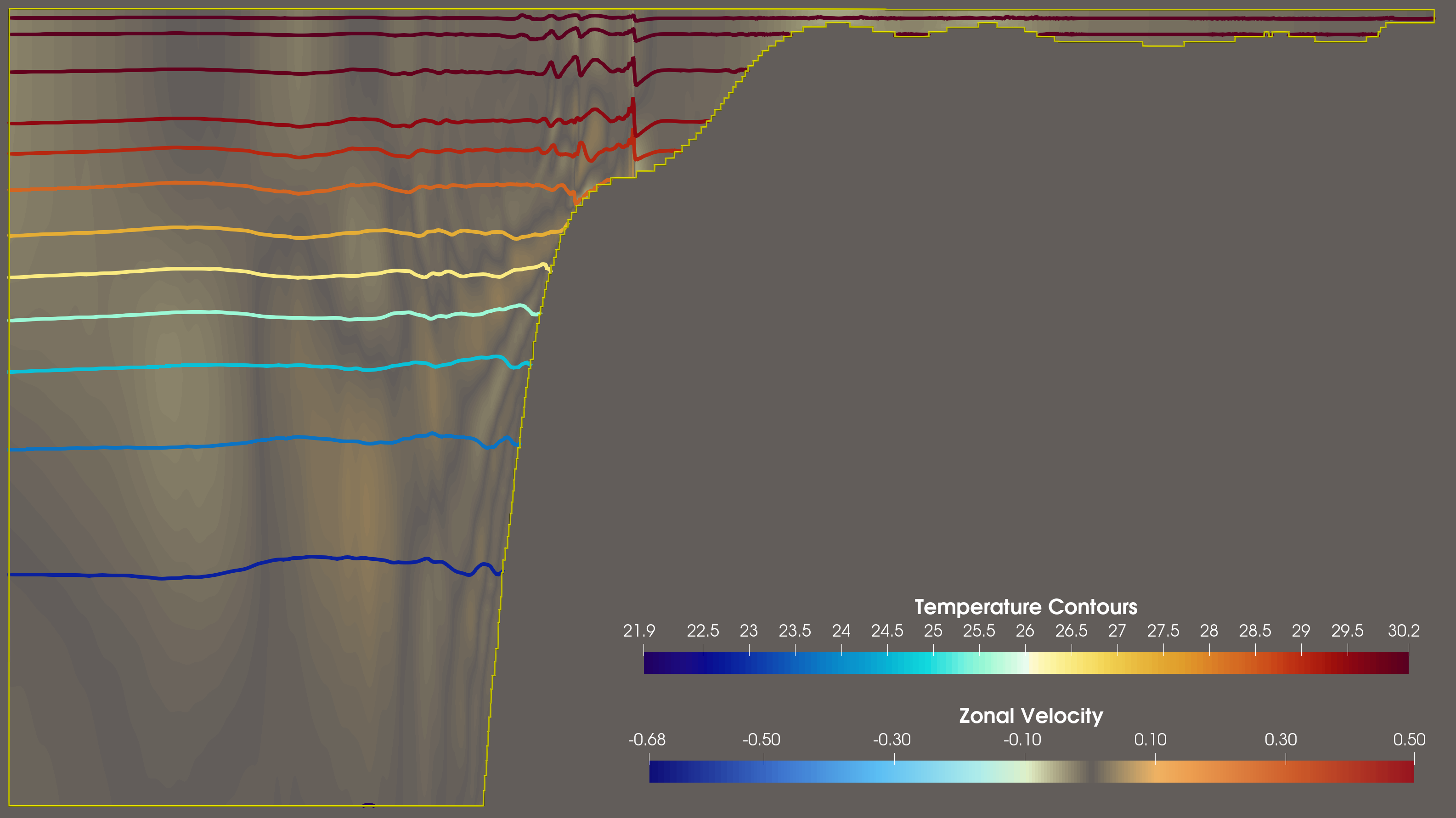}
    \caption{Nested high-resolution shoaling view, capturing fine-scale coastal interactions, energy dissipation, and nutrient mixing near the shoreline.}
    \label{fig:multiview}
\end{subfigure}
\hfill
\begin{subfigure}[t]{0.494\textwidth}
    \centering
    \includegraphics[trim=0 390px 0 0, clip, width=\linewidth]{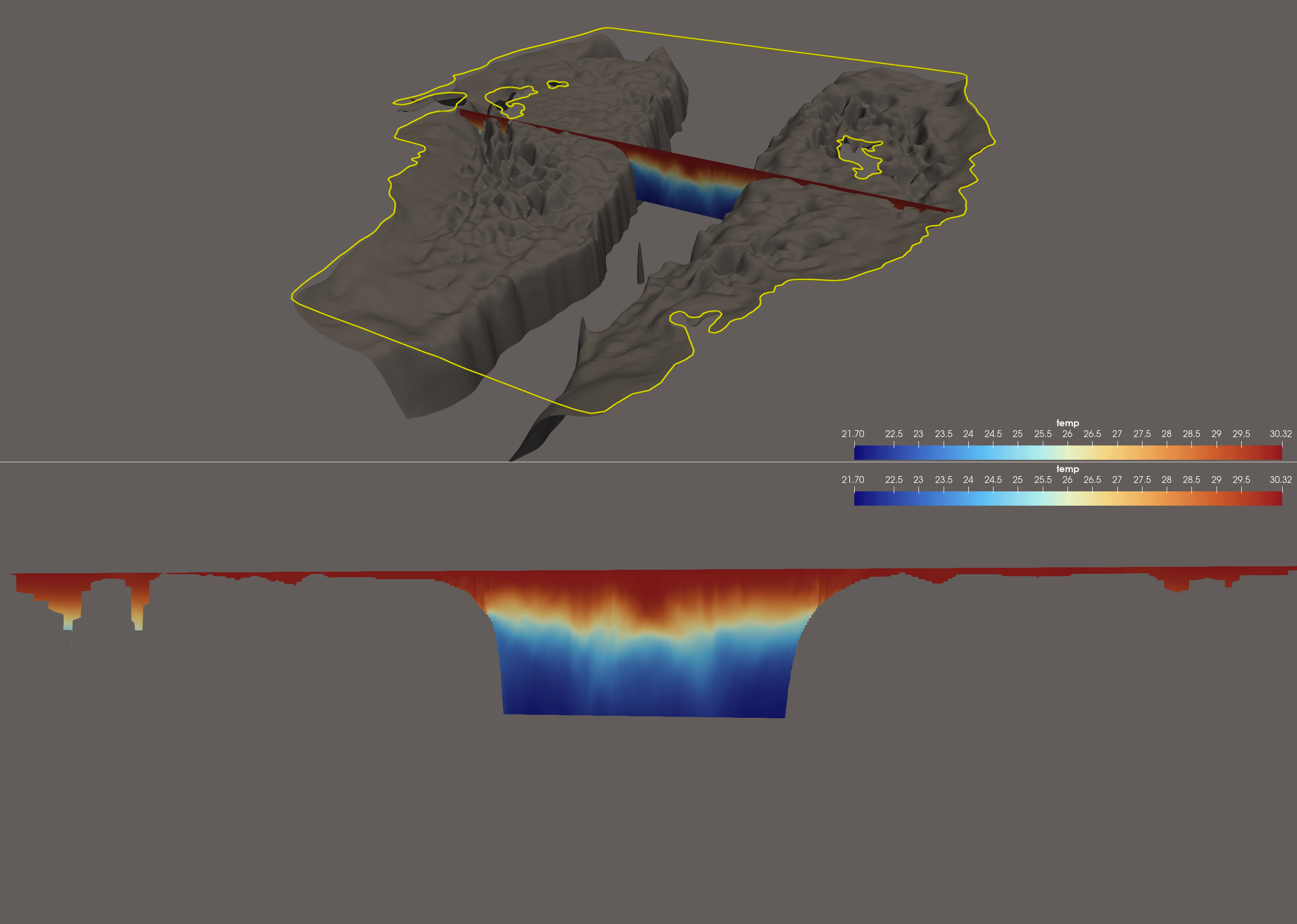}
    \caption{Interactive slice view displaying temperature (\texttt{temp}) gradients and density contours, allowing for detailed quantitative inspection of the water column.}
    \label{fig:interactive_temp}
\end{subfigure}

\caption{A coordinated, multi-view ocean data exploration dashboard at Time Step Jan 22, 2023, 18:00. The framework provides linked visualizations of surface tides (\ref{fig:eta}), bulk and density-driven flows (\ref{fig:barot}, \ref{fig:baroc}), isolated transverse waves (\ref{fig:cross_flow}), fine-scale coastal shoaling (\ref{fig:multiview}), and interactive 3D thermal cross-sections (\ref{fig:interactive_temp}). This coordinated approach enables analyzing complex flow dynamics from multiple perspectives simultaneously.}
\label{fig:figure-overviews}
\end{figure*}

\subsubsection*{Data and Variable Selection}
From the SUNTANS output, we selected vertical velocity (\texttt{w}) as the primary proxy for tracking internal waves. Supplementary variables include sea surface elevation (\texttt{eta}) for tidal context, and density (\texttt{rho}), temperature (\texttt{temp}),  and salinity (\texttt{salt}) for analyzing mixing. A derived \textbf{cross-flow vector} ($\vec{u}_{\perp}$) was conceptualized to isolate transverse wave features by filtering out the dominant tidal advection.

\subsubsection*{Multi-View Dashboard Composition}
To support coordinated analysis of flow-related variables without visual overload, the system is organized as a set of linked views (Fig.~\ref{fig:figure-overviews}), each targeting specific aspects of the task abstractions:

\begin{enumerate}
    \item \textbf{Main 3D Volumetric View (T2-T3) (Fig.~\ref{fig:teaser}):} 
    This view serves as the primary context window. It renders the 3D bathymetry (clipped below 200m to focus on the oceanic shelf) alongside a volume rendering of the vertical velocity (\texttt{w}). 
    
         \textit{\textbf{Visual Encoding:}} To reveal the internal wave structure without obscuring the domain, we designed a specific transparency transfer function that assigns total transparency to values near zero. This renders low-velocity regions invisible, emphasizing the significant vertical velocities associated with wave fronts. We utilized the "Fast" colormap, transitioning from blue (negative values) through white-yellow (near-zero) to red (positive values), to clearly distinguish between upwelling and downwelling currents.

    \item \textbf{Surface Context View (T2-T3) (Fig.~\ref{fig:eta}):} 
    This view visualizes the tidal context at the surface using the \texttt{eta} variable, providing contextual information about surface-driven dynamics.
    
    \textit{\textbf{Visual Encoding:}} The surface flow is represented by streamlines colored by sea surface elevation. Similar to the volume, transparency is applied to regions of minimal displacement, ensuring that only significant tidal variances are displayed.

    \item \textbf{Multiscale Flow Components View (Barotropic–Baroclinic) (T4): (Figs.~\ref{fig:barot} and \ref{fig:baroc}):} 
    Two dedicated views allow scientists to distinguish between the depth-averaged bulk motion (Barotropic) and the density-driven internal shear (Baroclinic).
    
        \textit{\textbf{Visual Encoding:}} To visualize these dense 2D vector fields without clutter, we utilized sparse streamlines to show continuity and glyphs to indicate direction. Crucially, we mapped opacity to \texttt{flow magnitude}. This acts as a visual filter, causing low-velocity regions to fade so the viewer can focus solely on high-energy transport pathways.

    {\color{black}\item \textbf{Cross-Flow View for Transverse Wave Detection (T4) (Fig.~\ref{fig:cross_flow}):} 
    This view is designed to reveal transverse wave-related motion that can be obscured by the dominant background flow. In multiscale ocean simulations, large-scale tidal advection often dominates vector-field visualizations, making it difficult to identify weaker but scientifically important lateral propagation patterns associated with internal waves. To address this, we derive a cross-flow vector field using a local reference direction defined from the dominant background flow, represented here by the barotropic velocity field. First, we compute a local unit reference direction by normalizing the \texttt{barotropic} velocity, $\hat{u}_{bt} = \vec{u}_{bt} / \|\vec{u}_{bt}\|$. We then project the \texttt{baroclinic} velocity, $\vec{u}_{bc}$, onto this local reference direction to obtain the component aligned with the dominant flow, $(\vec{u}_{bc} \cdot \hat{u}_{bt})\hat{u}_{bt}$. The transverse residual, or cross-flow component, is computed by subtracting this aligned component from the original \texttt{baroclinic} velocity: $\vec{u}_{\perp} = \vec{u}_{bc} - (\vec{u}_{bc} \cdot \hat{u}_{bt})\hat{u}_{bt}$.     
    The resulting field, $\vec{u}_{\perp}$, represents the portion of the baroclinic motion that is locally transverse to the dominant tidal advection and is visualized using the same streamline-and-glyph encoding as the other flow views for visual consistency.

Unlike classical approaches such as Helmholtz-Hodge decomposition \cite{helmholtz1858,schmid2001}, which separates vector fields into divergence-free and curl-free components, our formulation is directionally constrained by the dominant flow and separates the baroclinic motion into flow-aligned and transverse contributions. It also differs from conventional barotropic-baroclinic separation alone: after the baroclinic component is obtained, we further decompose it into a component aligned with the dominant barotropic flow and a transverse residual component. This makes the cross-flow view particularly useful for visually detecting transverse internal-wave propagation, lateral energy pathways, and wave-coast interactions that may remain hidden in raw velocity, barotropic, or baroclinic flow visualizations. Conventional flow separation techniques, such as vortex detection and shear-based analysis, primarily capture rotational structures or boundary-layer dynamics. However, they do not directly isolate motion transverse to a physically meaningful background-flow direction and may miss wave-driven lateral transport when it occurs within strong advective currents. In practice, this view provides a task-driven and physically interpretable representation that helps distinguish wave-driven transverse dynamics from the stronger background advective flow.}

    \item \textbf{Interactive Multivariate Inspection Views (T5):}  
    We enable multivariate reasoning through linked views and interactive slicing. To enable detailed quantitative analysis, the dashboard includes an Interactive Slice View (\textbf{Fig.~\ref{fig:interactive_temp}}) where users can manually position a plane to inspect temperature gradients and density (\texttt{rho}) contours. 
    In the initial design iterations, the domain scientists requested a view that could display high-resolution simulation output for a selected spatial area in the coastal regions to capture fine-scale shoaling effects~\cite{dean_dalrymple_1991}. The Shoaling View (\textbf{Fig.~\ref{fig:multiview}}) presents a 2D slice from a nested high-resolution simulation to capture fine-scale dissipation dynamics near the shore.    
\end{enumerate}

\subsection{Implementation of 3D Visualization Framework}

The framework is implemented in ParaView~6.0.1~\cite{ahrens2005paraview} using reproducible script-driven workflow consisting of preprocessing scripts, state files, and HPC runner scripts to operationalize the task-driven design (T1–T5). 
Hybrid-grid volumetric reconstruction and conversion of multiscale flow components, such as barotropic and baroclinic fluxes, into VTK-compatible meshes enable ParaView’s streamline-based analysis.
\textbf{Visualization Workflow}: 
The primary 3D view resamples the unstructured volume onto a ($1600 \times 1600 \times 64$) Cartesian grid for smooth wavefronts. 
Multiscale flows use \texttt{StreamTracer} and \texttt{Glyph} filters with magnitude-based opacity. Cross-flow components are computed on-the-fly using customized \texttt{Python scripts}, and surface context views extract the top layer with thresholded scalar mapping.
\textbf{Automation and HPC Deployment}: The workflow runs on Shaheen~III~\cite{shaheen3}, with Python scripts managing SLURM submission, ParaView executables, state files, and plugins. Rendering uses batch mode (\texttt{pvbatch}) with hybrid MPI+OpenMP (1 MPI × 128 threads) for efficient node-level utilization. Resulting image sequences are post-processed into MP4 videos with overlaid 2D plots onto the 3D visualizations.

\section{Case Studies}

We present three case studies demonstrating how the proposed visualization approach supports insight generation and analytical efficiency. We report domain-relevant findings identified by experts and contrast them with limitations of traditional slice-based workflows.

\subsection{Case Study 1: Generation and Evolution of NLIWs Flow}

\begin{figure*}[t!]
\centering
\includegraphics[width=\linewidth]{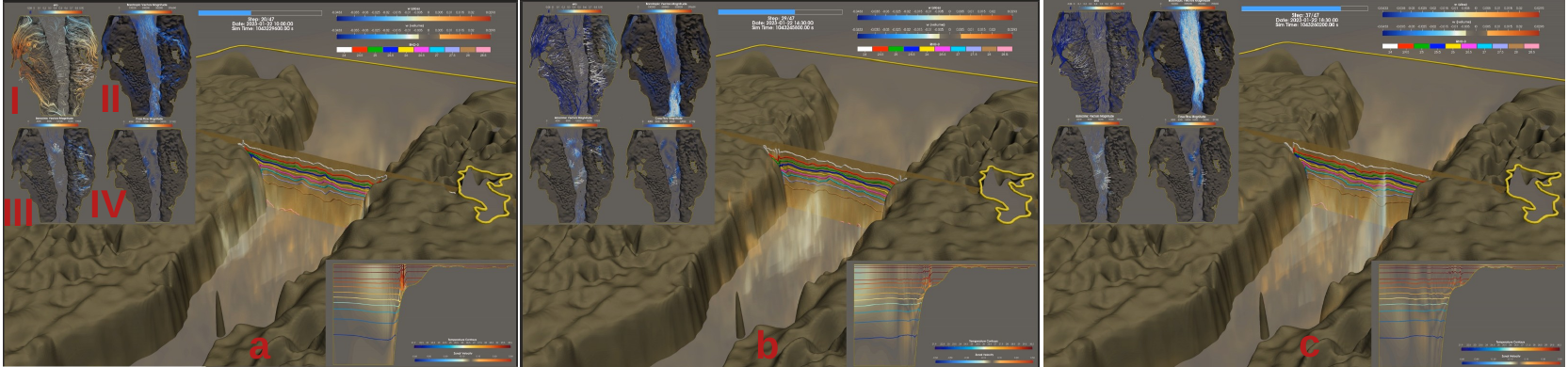}
\caption{Analyzing internal wave lifecycles and energy transport using the visualization framework. Panel I captures the 3D structural evolution of the waves across three time steps. Panels II, III, and IV detail the underlying energy mechanics, revealing the transition from barotropic tidal energy to baroclinic energy, along with isolated cross-flow transverse waves.}
\label{fig:case1}
\end{figure*}

Using the coordinated multi-view environment, domain experts analyzed the lifecycle of NLIWs across multiple time steps. The integration of volumetric rendering (vertical velocity), surface context (sea surface height), and synchronized transects enabled continuous tracking of wave generation, propagation, and transformation, directly supporting feature identification and spatiotemporal analysis (T3--T5).
Barotropic tidal flow over steep topography displaces {\color{black}isopycnals}, generating internal waves that initially propagate as linear waves but progressively steepen into nonlinear solitary-like waves under nonlinearity and nonhydrostatic dispersion.
The top-left inset (I) in Fig.~\ref{fig:case1} shows surface flow streamlines and sea surface height ($\eta$) overlaid on bathymetry. Across time steps, the surface flow diverges from the central region and crosses the continental shelf break toward shallower areas associated with elevated sea level.
Fig.~\ref{fig:case1}(b) captures the moment when these vertical disturbances propagate toward each other and converge near the central Red Sea. The vertical velocities (shading) highlight the north–south extent of the disturbance, revealing its fully three-dimensional structure. Fig.~\ref{fig:case1}(c) shows the stage when the eastward- and westward-propagating waves grow in amplitude and begin to disintegrate into solitary-like waves$-$a single steep wave pulse that travels mostly on its own and slowly changes shape and loses energy.
Taken together, the evolution of isopycnals along the transect, the vertical velocity fields, and the surface flow directions provide a comprehensive view of the generation regions and subsequent evolution of NLIWs.

\textit{\textbf{Finding:}This case study demonstrates how the framework enables direct visual analysis of NLIW dynamics through integrated 3D representations and linked multivariate views. In contrast to traditional workflows based on sequential 2D transects, which require manual spatial and temporal alignment, coordinated visualization reduces interpretive ambiguity. Experts were able to identify wave convergence zones and track propagation patterns more efficiently than with traditional slice-based workflows, which required inspecting multiple transects and manually aligning features across time steps. Domain experts reported that 3D views improved their ability to perceive spatial coherence, track wave propagation, and identify structural transformations without reconstructing mental models from multiple plots.}

\subsection{Case Study 2: Barotropic and Baroclinic Tidal Energy} 

We analyzed energy transport by visualizing barotropic flow, baroclinic flux, and the derived cross-flow field in synchronized views, enabling coordinated multivariate analysis (T4--T5). Quantifying the energy of NLIWs is essential for understanding the conversion of barotropic tidal energy into internal waves, their subsequent transport, and their dissipation.
The depth-integrated barotropic energy flux characterizes the magnitude and direction of available tidal energy. The top-right inset (II) in Fig.~\ref{fig:case1} shows this flux across different tidal phases, revealing that barotropic energy predominantly propagates in the north--south direction along the central trench of the southern Red Sea.
This flux also represents the energy transferred from barotropic tides into internal waves. The arrows in the bottom-left inset (III) (Fig.~\ref{fig:case1}) indicate the direction of baroclinic energy propagation and highlight regions of active energy conversion. The cross-flow field (Section~\ref{sec:vis_design}), shown in the bottom-right inset (IV) (Fig.~\ref{fig:case1}), isolates energy components associated with NLIWs propagating in the east--west direction.

\textit{\textbf{Finding:} This case study demonstrates how the framework reveals energy transport pathways and the transition from barotropic to baroclinic energy across the Red Sea. Coordinated views enable simultaneous analysis of interacting flow components, while the cross-flow representation isolates wave-driven energy transport. Experts identified distinct conversion regions and lateral propagation patterns that are not evident in individual velocity or flux fields; in particular, the cross-flow view exposes transverse energy structures that remain hidden in standard visualizations. Compared to traditional workflows that require manual inspection and alignment of multiple vector fields, coordinated visualization enables more efficient isolation of wave-driven energy transport and improves interpretability.}

\begin{figure*}[t!]
\centering
\includegraphics[width=\linewidth]{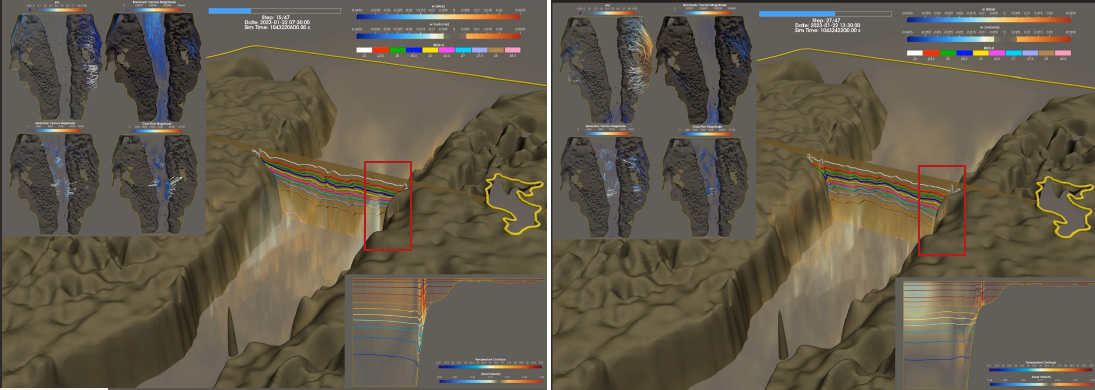}
\caption{Case Study 3: High-resolution analysis of internal wave shoaling and coastal interaction. The main view shows the wave's arrival at the western boundary of the coastal domain (indicated by the red rectangle). The detailed inset captures the subsequent shoaling phase, revealing a sharp upward displacement of isotherms as the wave spills over the shelf. This visualizes how cooler waters from depths greater than 60~m are forced into the 30–40~m range, a key mechanism for coastal nutrient transport.}
\label{fig:case3}
\end{figure*}
\vspace{-6pt}

\subsection{Case Study 3: Shoaling and Coastal Interaction} 
Solitary-like internal waves are known to travel long distances with minimal dissipation as they propagate through the ocean and interact with other dynamical processes. However, as these waves approach shallow regions, they shoal (the process by which waves move from deep water into shallow water) and dissipate, generating turbulence. During shoaling, low-temperature, nutrient-rich waters are lifted onto continental shelves, providing cooling and nutrient supply that can benefit coral reefs and influence local biological productivity.

Because shoaling and dissipation require even higher spatial and temporal resolution to be accurately captured, a two-dimensional (x–z) numerical simulation was carried out using finer horizontal (30 m) and vertical (1 m) grid spacings. Smoothed density and velocity fields from the 3D simulations were used as the initial and boundary conditions for the simulations. Fig.~\ref{fig:case3} shows two stages of wave evolution: the arrival of the wave at the western boundary of the high-resolution domain (indicated by the red rectangle), and the subsequent shoaling phase. Notably, during shoaling, the isotherms exhibit a sharp upward displacement (as shown in the Shoaling View in the bottom-right inset of Fig.~\ref{fig:case3}), with cooler waters originating from depths greater than 60 m intruding into the 30–40 m depth range. 
This reveals potential nutrient transport in shallow Red Sea regions, supporting analysis of marine ecosystem impacts, wave–structure interaction, and wave–coast processes.

Using the framework, experts analyzed shoaling through coordinated volumetric context, transects, and localized high-resolution views, linking large-scale wave structure to fine-scale coastal processes. The hybrid-grid reconstruction (T1) enables consistent integration of 3D and 2D simulation data, while the bathymetric context supports spatial interpretation (T2). Interactive slicing and localized views facilitate tracking of wave deformation and shoaling progression (T3), and coordinated multivariate visualization of temperature, velocity, and density supports analysis of interacting physical processes (T4--T5).

\textit{\textbf{Finding:} This case study demonstrates how the framework enables detailed analysis of shoaling dynamics and coastal interaction. Experts observed sharp upward displacement of isotherms, revealing the transport of cooler, nutrient-rich water into shallow regions. Coordinated views allowed them to correlate wave structure, temperature gradients, and bathymetry, providing a clearer understanding of the mechanisms driving coastal mixing. Compared to conventional workflows requiring multiple high-resolution slices and manual cross-referencing of variables, the interactive and linked visualization enabled more efficient identification of shoaling-induced mixing patterns and improved interpretability of fine-scale dynamics.}

\textit{\textbf{Case Studies Summary:} Across the three case studies, the framework enables domain experts to interactively analyze multiscale and multivariate ocean phenomena through coordinated, synchronized views. Presenting complementary representations interactively and side by side supports direct comparison, exploration, and correlation of features across space, time, and variables. This integrated approach enables more efficient analysis and reduces reliance on manual, slice-based workflows. In particular, the framework enables insights that are difficult or not feasible with conventional methods, including the identification of transverse energy pathways via the cross-flow view, direct tracking of three-dimensional wave convergence using volumetric and linked views, and analysis of shoaling-induced mixing patterns through coordinated multivariate visualization.}

\section{Evaluation with Domain Experts}
We evaluate our approach through an insight-based qualitative study with domain experts, focusing on its effectiveness in supporting the task abstractions (T1–T5) and enabling new scientific understanding compared to traditional slice-based workflows.

\subsection{Study Design and Methodology}
\textbf{Participants:} We conducted evaluation sessions with five domain experts (E1-E5) specializing in physical oceanography, ocean modeling, forecasting, and coral reef bleaching. All experts posses over a decade of advanced research experience and are familiar with ocean flow dynamics and physical oceanography. They have extensive experience analyzing ocean simulation data using conventional tools such as MATLAB, Python, and domain-specific software (e.g., Ferret, CDO, NCO, and GrADS), along with 2D slice-based workflows. These experts are not co-authors of this paper and served as independent evaluators. 
All domain experts participated voluntarily and provided informed consent before participating. Their feedback was reported in anonymized form, and no sensitive personal information was collected. 

\textbf{Procedure and Study Design:} The evaluation was structured as a 60-minute synchronous observation and interview session. The study was divided into three distinct phases:

    \textit{\textbf{Context and Tutorial (15 minutes):}} 
    We first explained the ocean flow model and associated simulation dataset,  including the spatial domain, simulation period, resolution, and bathymetry details. We then introduced the visualization workflow, detailing each component, analysis goals, and the system interface.
        
    \textit{\textbf{Exploratory Analysis (30 minutes):}} We demonstrated the full system and conducted guided analysis sessions. The experts were given the opportunity to direct the exploration, requesting specific analyses of their choosing while asking clarification questions about specific components and features. The analysis sessions involved investigating wave dynamics, dominant flows and directional contributions, energy transport, and coastal processes, while verbalizing observations.  
    
    \textit{\textbf{Guided Interview (15 minutes):}} 
    Semi-structured discussion on analytical capabilities, comparison with existing workflows, and perceive strengths and limitations.

\textbf{Data Analysis Strategy:} During the exploratory phase, we encouraged the experts to actively vocalize their hypotheses and observations as they interacted with the 3D environment. Detailed observational notes and session feedback were recorded. Following the session, we distilled the experts' insights into core themes regarding visual effectiveness and domain utility.

\begin{table}[t]
\captionsetup{skip=2pt, font=footnotesize}
\centering
\caption{Domain Experts' Insights Summary}
\label{tab:insights}
\small
\setlength{\tabcolsep}{4pt}
\renewcommand{\arraystretch}{1.1}

\begin{tabularx}{\linewidth}{|X|X|>{\centering\arraybackslash}p{2cm}|}
\hline
\multicolumn{1}{|c|}{\textbf{Insights}} & 
\multicolumn{1}{c|}{\textbf{Enabled By}} & 
\multicolumn{1}{c|}{\textbf{Prior Capability}} \\ 
\hline

Detection of transverse wave propagation patterns 
& Cross-flow decomposition 
& No \\ 
\hline

Identification of cross-basin energy transport pathways and convergence zones 
& Linked multi-view analysis 
& Difficult \\ 
\hline

Continuous tracking of wave lifecycle (generation → propagation → dissipation) in 3D 
& Volumetric rendering, temporal navigation 
& No \\ 
\hline

Correlation of flow dynamics with thermodynamic variables (temperature, salinity, density) 
& Linked multivariate views 
& Difficult \\ 
\hline

Identification of shoaling-induced vertical transport of cooler, nutrient-rich waters 
& Interactive slicing + high-resolution views 
& No \\ 
\hline

\end{tabularx}

\end{table}

\subsection{Expert Feedback and Domain Insights}
The feedback and observations from domain experts were broadly focused on the different task abstractions (T1-T5). 
We summarize key insights and workflow improvements identified during expert evaluation in Tables~\ref{tab:insights} and~\ref{tab:workflow}. Table~\ref{tab:insights} presents the insights captured during domain expert evaluation, the features that enabled them, and their feasibility in traditional workflows, highlighting how the proposed framework facilitates new insights and simplifies complex analyses. Table~\ref{tab:workflow} compares traditional workflows with the proposed framework across common analysis tasks, emphasizing gains in efficiency, usability, and time to insight. These findings are based on discussions with domain experts during the exploratory analysis and interview phase.

\textbf{Visual Effectiveness and Flow Decomposition:} 
All the experts praised the specialized 2D views. E1 inquired about the available threshold configuration options and appreciated the ability to interactively adjust the transfer functions and associated color encodings. E1 noted that the surface context view effectively highlighted the presence of shallow regions. Regarding the flow decomposition and multiscale dynamics, E1 commented that the barotropic view provided clear insights into overall tidal patterns, while the baroclinic view facilitated the identification of regions with strong magnitudes. E1 specifically noted that the cross-flow view is highly useful, as it \textit{"allows one to clearly see the propagation of high energies."} E2 echoed this sentiment, stating that while the baroclinic view helps identify complicated regions, the cross-flow view is even clearer, making wave fronts and crests easily identifiable. Commenting on the barotropic view, E2 remarked, \textit{“This is cool, it describes how the main tidal currents are moving... I would love to use such an interactive tool in my daily workflow.”} E2 also highlighted the usefulness of the streamlines, particularly within the strait. E3 suggested that incorporating tidal phases (high and low tides) alongside barotropic and baroclinic flows could provide additional insight into the observed flow patterns. E3 commented that all information related to dominant and transverse flow, as well as interactions with coastal regions, is presented within a cohesive multiview environment. E4 enquired about the capabilities to inject custom functionality in the processing pipelines, and also the capabilities to interactively probe different sections of the flow fields.

\textbf{3D Contextualization and Shoaling Dynamics:} The 3D volumetric rendering was a major highlight for the evaluators. E1 praised the 3D view, noting, \textit{“This particular animation is very nice; propagation and dissipation are very easy to follow, and this is quite difficult to capture in 2D slices.”} E1 mentioned that the strongest waves could easily be identified using the vertical velocities. Furthermore, E1 appreciated the high-fidelity simulation for capturing shoaling details, noting how easily one can observe the waves propagating, diffusing, and transferring energy after interacting with the coast. E2 specifically liked the 3D view in the context of shoaling, stating, \textit{“3D is better and has more dynamics. It is quite important for the ecosystem to see how it goes and brings in nutrients.”} E5 particularly appreciated the interactive multivariate inspection capabilities and the Shoaling view, and noted that \textit{“This kind of environment could be useful in understanding the accumulation of phytoplankton, correlations with ocean dynamics, and other contributing factors.”} E4 appreciated the ability to visualize internal tidal beams alongside Shoaling dynamics.  

\textbf{Workflow Integration and Future Improvements:} Overall, all participants appreciated the comprehensive suite of visualization tools and interactions. E1 praised the interaction techniques, particularly the ability to move the 2D slice within the 3D view and overlay density contours. E1 noted that the 3D interactions provided by this workflow could help validate hypotheses that are extremely difficult to test using their traditional 2D plotting tools. E2, E3, E4 similarly noted that this level of exploration and interactivity is currently unavailable in their standard domain tools. 
The experts also provided suggestions for future iterations. E1 suggested that time information could be displayed more prominently and proposed overlaying multiple variables in different views (e.g., sea surface height with temperature). E2 suggested filtering out the sea surface height displayed in shallow regions, noting that it can sometimes overpower the view. E2 and E4 also requested the ability to define specific locations of interest within the spatial domain, plot wave amplitude spatially to track propagation, and release 3D particles to trace nutrient transport pathways relevant to fisheries (E2, E5). E5 suggested that overlaying additional information (e.g., coral maps) could be useful in studying marine biology or coral reef ecology.

\begin{table}[t]
\captionsetup{skip=2pt, font=footnotesize}
\centering

\caption{Workflow Comparison}
\label{tab:workflow}

\small
\setlength{\tabcolsep}{4pt}
\renewcommand{\arraystretch}{1.1}

\begin{tabularx}{\linewidth}{|X|X|>{\centering\arraybackslash}p{3cm}|}
\hline
\multicolumn{1}{|c|}{\textbf{Task Description}} & 
\multicolumn{1}{c|}{\textbf{Traditional Workflow}} & 
\multicolumn{1}{c|}{\textbf{Proposed Framework}} \\ 
\hline

Extract spatial transects 
& Manual scripting and repeated plotting 
& Interactive selection and slicing \\ 
\hline

Multi-variable analysis 
& Sequential inspection of separate plots 
& Simultaneous linked multi-view exploration \\ 
\hline

Flow component analysis 
& Indirect inference or manual filtering 
& Direct decomposition (cross-flow view) \\ 
\hline

Time to insight 
& $\sim$30--60 minutes (iterative) 
& $\sim$5--10 minutes (interactive exploration) \\ 
\hline

Cognitive effort 
& High (Manual alignment and interpretation) 
& Reduced (Integrated spatial and variable context) \\ 
\hline

\end{tabularx}

\end{table}

Our evaluation is limited to a small number of domain experts, as engaging large numbers of highly specialized experts is often challenging due to their availability and expertise requirements. Although the approach is designed to be generalizable, further validation across different domains and datasets is required.

\section{Discussion and Lessons Learned}
We discuss the implications, limitations, and generalizability of our approach, with respect to the task abstractions (T1–T5) and the insights derived from case studies and evaluation. 

\subsection{Scientific Impact and Domain Feedback}
Our approach demonstrates that integrating hybrid-grid volumetric reconstruction (T1), context-aware visualization (T2), feature tracking (T3), flow decomposition (T4), and multivariate exploration (T5) enables domain experts to uncover complex ocean flow dynamics more effectively than traditional slice-based workflows. In particular, the cross-flow decomposition and coordinated multi-view design were critical in revealing transverse wave propagation and energy transport pathways that were previously difficult to observe.

The most significant success of this framework, as validated by domain experts, is the ability to visualize the vertical water velocity (\texttt{w}) in a fully volumetric 3D context. Prior to this work, scientists primarily relied on 2D planar cross-sections or depth-averaged 2D visualizations. During the evaluation, experts noted that the ability to see the complete 3D structure of \texttt{w} was powerful, enabling them to track the formation and complex evolution of internal waves in ways that were previously impossible with their standard toolsets. This visual validation confirmed hypotheses regarding the disintegration of waves into solitary-like structures (Case Study 1) and provided a new method for verifying model fidelity against physical expectations.

\subsection{Visualization Design and Implementation}

Visualizing complex ocean datasets requires sophisticated, custom workflows. Raw oceanographic data must first be cleaned and geometrically transformed before rendering. Our framework, built on ParaView, implements customized workflows for multiscale, multivariate datasets on complex hybrid grids, and is generalizable to similar scientific visualization platforms. The domain-specific functionality we developed is not readily available in standard SciVis libraries.
Visual representations rely on extensive custom pipelines within ParaView, including tailored transparency transfer functions and specialized data extraction methods to highlight key physical phenomena. Without these pipelines, raw volumes appear as opaque blocks, which limits usability for domain scientists without significant technical expertise.

Our custom volume reconstruction method addresses a common challenge in ocean modeling by extruding horizontally unstructured 2D polygons along varying vertical coordinates to generate cohesive wedges and hexahedra. This approach is model-agnostic and can be applied to other complex volumetric datasets with hybrid grids, resolving a frequent bottleneck in coastal modeling simulations.

\subsection{Failure Cases and Limitations }
Despite its effectiveness, the approach has several limitations: 

\textbf{Sensitivity to Weak Dominant Flow (T4)}: The cross-flow decomposition relies on a well-defined dominant flow direction. In regions where the barotropic velocity magnitude is very low or highly variable, the decomposition may become unstable or less meaningful, potentially introducing noisy or ambiguous transverse components.  

\textbf{Occlusion and Transfer Function Sensitivity (T3)}: Volumetric rendering depends on transfer function design. While magnitude-based filtering reduces clutter, inappropriate parameter choices may suppress subtle but relevant structures or introduce visual bias.  

\textbf{Resolution and Data Fidelity (T1)}. The resampling step used for volumetric rendering may smooth fine-scale features, particularly in highly localized or turbulent regions. This introduces a trade-off between visual clarity and data fidelity.

\subsection{Scalability and Performance}
Scalability is critical given the size of ocean datasets. \textbf{Data Size and Processing (T1)}: Hybrid-grid reconstruction and cell-to-point interpolation add preprocessing overhead but are performed once per dataset, enabling downstream interactive analysis. \textbf{Rendering Performance (T3–T5)}: Volume Rendering of large unstructured grids is computationally expensive; structured resampling improves performance and reduces artifacts, with a trade-off between resolution and interactivity. \textbf{HPC Integration}: The system supports HPC environments for batch processing and rendering, and single-node, multi-threaded execution provides a balance between performance and visual quality, while further optimizations such as distributed rendering remain future work.

\section{Conclusion and Future Work}
We presented a task-driven framework for multiscale, multivariate flow data on hybrid grids that combines volumetric reconstruction, feature tracking, cross-flow decomposition, and coordinated multiview exploration with magnitude-based encodings. Three case studies and expert feedback demonstrate improved interpretation of 3D internal-wave dynamics, including transverse energy pathways and coastal shoaling. Future work includes MPI-based reconstruction, in situ visualization ~\cite{childs2020terminology}, a Trame-based web interface~\cite{Trame}, Adaptive Mesh Refinement (AMR)~\cite{BERGER198964,AMRex}, and open-source release of the reconstruction plugin.


\bibliographystyle{abbrv-doi-hyperref}

\acknowledgments{
 This work was supported by the Office of the Vice President for Research at King Abdullah University of Science and Technology (KAUST) through the Digital Arabian Peninsula Initiative (\#REP/1/7475-01-01).
}

\bibliography{template}

\clearpage
\appendix

\section{{\color{black}Appendix: Data Preprocessing and Modular Visualization Pipelines}}

The multiscale flow visualizations presented in this study rely on a custom data transformation and rendering pipeline. Because the numerical model outputs data on a complex hybrid grid (unstructured horizontal polygons with terrain-following vertical layers), standard visualization tools cannot natively interpret the vertical connectivity.  

To resolve this, we developed a custom \texttt{SuntansReader} plugin for ParaView that explicitly connects corresponding nodes across layers using the model’s depth coordinates, creating a cohesive 3D unstructured volume.  
We are currently investigating the public release of this plugin to facilitate broader community reuse.

\subsection*{Data Preprocessing and Range Extraction}
To ensure stability and accurate color mapping within the 3D environment, raw simulation outputs undergo a series of preprocessing steps:
\begin{itemize}
    \item \textbf{Data Cleaning:} A preprocessing script (\texttt{clean\_data.py}) handles invalid data by detecting numeric \texttt{\_FillValue} attributes and replacing values exceeding the fill threshold with \texttt{NaN}.  
    The script processes data in 512 MB chunks and converts integer arrays to 32-bit floats when necessary to prevent rendering artifacts.  
    \item \textbf{Global Data Bounds:} To establish consistent visual encodings, custom routines automate the extraction of global minimum, maximum, and second-highest values across all timesteps for target variables.  
    \item \textbf{2D Spatial Interpolation:} For supplementary planar analyses, spatial scripts utilize Inverse Distance Weighting (IDW) via \texttt{cKDTree} to project unstructured hybrid grid data onto high-resolution 2D query paths, calculating derived perturbations and animating vertical slices.  
\end{itemize}

\subsection*{Core Prerequisite: Spatial Transformation \& Optimization}
Before generating individual 3D views, the data must be geometrically normalized:
\begin{itemize}
    \item \textbf{Mesh Extraction \& Appending:} The custom \texttt{SuntansReader} loads the grid, and \textbf{Extract Block} filters isolate the native \texttt{r} and \texttt{w} mesh topologies.  
    An \textbf{Append Attributes} filter merges these blocks to enable cross-mesh variable computation.  
    \item \textbf{Vertical Exaggeration:} To compensate for the severe aspect ratio of the Red Sea basin, a \textbf{Transform} is applied to the appended mesh.  
    The geometry is translated vertically via $\text{Translate} = [0.0, 0.0, -25.0]$ and depth-exaggerated using $\text{Scale} = [1.0, 1.0, -200.0]$.  
\end{itemize}

\subsection*{1. Main 3D Volumetric View (Vertical Velocity \& Density)}
This pipeline renders the full 3D spatial context of internal wave fronts and upwelling/downwelling currents.  
\begin{itemize}
    \item \textbf{Resampling:} Apply a \textbf{Resample To Image} filter to the transformed mesh with $\text{SamplingDimensions} = [500, 500, 46]$ to generate a regular grid optimized for volume rendering.  
    \item \textbf{Volume Rendering Settings:} Map the point-centered vertical velocity array (\texttt{w}) to a ``Fast'' diverging colormap (blue to white-yellow to red).  
    Apply a custom piecewise opacity function to set opacity to $0.0$ for near-zero ambient velocities, eliminating background clutter.  
    \item \textbf{Density Conversion \& Isosurfaces:} Apply a \textbf{Calculator} filter ($\text{rho} \times 1000$) to convert normalized density to physical values ($\text{kg/m}^3$).  
    Connect a \textbf{Contour} filter to define explicit isosurfaces at $0.5$ intervals ranging from $24.0$ to $29.5$.  
    \item \textbf{Bathymetry:} Render the 3D bathymetry geometry, clipping it below a 200m depth threshold to focus viewer attention on the oceanic shelf dynamics.  
\end{itemize}

\subsection*{2. Surface Context View (Eta)}
This pipeline isolates sea surface height anomalies as an indicator of tidal waves.  
\begin{itemize}
    \item \textbf{Surface Extraction:} Apply a \textbf{Slice} filter positioned at the top of the transformed domain to extract the sea surface boundary.
    \item \textbf{Flow Encodings:} Attach a \textbf{Stream Tracer} utilizing the horizontal velocity fields (\texttt{uc}, \texttt{vc}) and pass the streamlines into a \textbf{Glyph} filter for vector direction.  
    Color the resulting streams and glyphs by sea surface elevation (\texttt{eta}).  
    \item \textbf{Filtering:} Apply magnitude-based transparency so regions with minimal displacement fade out entirely.  
\end{itemize}

\subsection*{3. Barotropic View (Depth-Averaged Bulk Flow)}
This pipeline displays the macro-scale bulk motion of the water column.  
\begin{itemize}
    \item \textbf{Flow Tracing:} Pass the depth-integrated velocity flux array into a \textbf{Stream Tracer} configured to integrate across the 2D plane.  
    \item \textbf{Glyph Mapping:} Apply a \textbf{Glyph} filter configured with directional arrows.  
    Map the opacity of both the streamlines and glyphs directly to the velocity magnitude to visually suppress low-energy current regions.  
\end{itemize}

\subsection*{4. Baroclinic View (Internal Density Shear)}
This pipeline captures internal shear waves and vertical energy beams.  
\begin{itemize}
    \item \textbf{Flux Extraction:} Isolate the baroclinic flux vector field.  
    \item \textbf{Vector Field Mapping:} Use an identical visual pipeline to the Barotropic view (sparse \textbf{Stream Tracer} and \textbf{Glyph} sequence).  
    Opacity mapped to velocity magnitude automatically exposes high-energy internal wave paths.  
\end{itemize}

\subsection*{5. Cross-Flow View (Mathematically Isolated Transverse Waves)}
To isolate subtle transverse wave dynamics masking within dominant currents, we implement a directionally constrained cross-flow vector field calculation directly within a \textbf{Programmable Filter}:  

$$ \hat{u}_{bt} = \frac{\vec{u}_{bt}}{\|\vec{u}_{bt}\|} $$

$$ \vec{u}_{\parallel} = (\vec{u}_{bc} \cdot \hat{u}_{bt})\hat{u}_{bt} $$

$$ \vec{u}_{\perp} = \vec{u}_{bc} - \vec{u}_{\parallel} $$

\begin{itemize}
    \item \textbf{Mathematical Step:} The filter normalizes the barotropic bulk velocity vector ($\vec{u}_{bt}$) to establish a local directional unit vector ($\hat{u}_{bt}$).  
    The baroclinic velocity ($\vec{u}_{bc}$) is projected onto this unit vector to isolate the parallel flow component ($\vec{u}_{\parallel}$).  
    Subtracting this yields the isolated cross-flow vector field ($\vec{u}_{\perp}$).  
    \item \textbf{Rendering:} Inject the custom $\vec{u}_{\perp}$ array into a standard \textbf{Stream Tracer} and \textbf{Glyph} pipeline with magnitude-mapped opacity to isolate east-west wave propagation.  
\end{itemize}

\subsection*{6. Shoaling View (Nested Fine-Scale Domain)}
This pipeline tracks fine-scale coastal boundaries and wave breaking.  
\begin{itemize}
    \item \textbf{High-Res Ingestion:} Load the specialized high-resolution 2D transect dataset featuring 30m horizontal and 1m vertical spacing. 
    \item \textbf{Isotherm Tracking:} Use a \textbf{Contour} filter tied to the high-resolution temperature (\texttt{temp}) array to map sharp upward displacements of cooler waters originating from $>60$m depth being forced into the $30$--$40$m depth range over the shallow shelf. 
\end{itemize}

\subsection*{7. Interactive Slice View (Thermodynamic Probing)}
This pipeline allows interactive quantitative evaluation along user-defined sections.  
\begin{itemize}
    \item \textbf{Planar Slicing:} Create an interactive \textbf{Slice} filter with the spatial widget enabled. 
    Map the point-centered temperature (\texttt{temp}) variable across the continuous slice surface. 
    \item \textbf{Density Overlay:} Attach a second \textbf{Contour} directly to the slice data, calculating density boundaries at $0.5$ intervals to draw sharp isopycnal lines over the temperature profile.
    \item \textbf{Temporal Context:} Attach a \textbf{Time Step Progress Bar} and a Python-based \textbf{Programmable Annotation} filter to extract the current simulation time and format it as a persistent Date/Time string overlaid on the dashboard.  
\end{itemize}

\end{document}